\documentclass[prd, nofootinbib, floatfix, notitlepage]{revtex4-1}

\usepackage{amsmath,amsfonts,amssymb}
\usepackage{mathrsfs}
\usepackage{graphicx}
\usepackage[english]{babel} 
\usepackage{hyperref} 
\usepackage{color}
\usepackage{bbold}
\usepackage{adjustbox}

\newcommand{\be}{\begin{equation}}
\newcommand{\ee}{\end{equation}}
\newcommand{\bes}{\begin{equation*}}
\newcommand{\ees}{\end{equation*}}

  \usepackage[utf8]{inputenc} 

\def\VEV#1{\left\langle #1 \right\rangle}
\newcommand{\wigner}[6]{ \begin{pmatrix}
  #1 & #2 & #3 \\
  #4 & #5 & #6 
\end{pmatrix}}

\newcommand{\jcap}{JCAP.}

\def\wigner#1#2#3#4#5#6{ \left( \begin{array}{ccc} #1 & #3 & #5
\\ #2 & #4 & #6 \\ \end{array} \right)}

\begin{document}

\title{Baryons still trace dark matter: probing CMB lensing maps for hidden isocurvature}

\author{Tristan L.~Smith$^1$}
\author{Julian B.~Mu\~noz$^2$}
\author{Rhiannon Smith$^1$}
\thanks{These authors contributed equally to this work.}
\author{Kyle Yee$^1$} \thanks{These authors contributed equally to this work.}
\author{Daniel Grin$^3$}
\affiliation{$^1$Department of Physics and Astronomy, Swarthmore College, 500 College Ave., Swarthmore, PA 19081, United States}
\affiliation{$^2$Department of Physics and Astronomy, Johns
				Hopkins University, 3400 N.\ Charles St., Baltimore, MD 21218, United States}
\affiliation{$^3$Department of Physics and Astronomy, Haverford College, 370 Lancaster Avenue, Haverford, PA 19041, United States}

\date{\today}

\begin{abstract}
Compensated isocurvature perturbations (CIPs) are primordial fluctuations that balance baryon and dark-matter isocurvature to leave the total matter density unperturbed. 
The effects of CIPs on the cosmic microwave background (CMB) anisotropies are similar to those produced by weak lensing of the CMB: smoothing of the power spectrum, and generation of non-Gaussian features. Previous work considered the CIP effects on the CMB power-spectrum but neglected to include the CIP effects on estimates of the lensing potential power spectrum (though its contribution to the non-Gaussian, connected, part of the CMB trispectrum).
Here, the CIP contribution to the standard estimator for the lensing potential power-spectrum is derived, and along with the CIP contributions to the CMB power-spectrum,  \textit{Planck} data is used to place limits on the root-mean-square CIP fluctuations on CMB scales, $\Delta_{\rm rms}^2(R_{\rm CMB})$. The resulting constraint of $\Delta_{\rm rms}^2(R_{\rm CMB}) < 4.3 \times 10^{-3}$ using this new technique improves on past work by a factor of $\sim 3$. We find that for \textit{Planck} data our constraints almost reach the sensitivity of the optimal CIP estimator.  The method presented here is currently the most sensitive probe of the amplitude of a scale-invariant CIP power spectrum placing an upper limit of $A_{\rm CIP}< 0.017$ at 95\% CL. Future measurements of the large-scale CMB lensing potential power spectrum could probe CIP amplitudes as low as $\Delta_{\rm rms}^2(R_{\rm CMB}) = 8 \times 10^{-5}$ ($A_{\rm CIP} = 3.2 \times 10^{-4}$). 
\end{abstract}

\maketitle

\section{Introduction}
\label{sec:intro}

The success of the standard cosmological model has been established using a wide range of observations: estimates of the primordial light element abundances predicted by standard big bang nucleosynthesis (BBN) \cite{PhysRev.73.803,RevModPhys.88.015004}, the observed isotropy and structure of the acoustic peaks in the cosmic microwave background (CMB) \cite{PhysRevLett.100.191302,PhysRevLett.101.011301,1992ApJ...396L...1S,2013ApJS..208...20B,Aghanim:2015xee}, increasingly restrictive upper limits to the level of non-gaussianity in the CMB \cite{Ade:2015ava}, the large-scale clustering of galaxies \cite{Alam:2016hwk}, and upper limits to non-standard initial conditions \cite{Bean:2006qz, Ade:2015lrj}--just to name a few.  In addition to these successes, there are some inconsistencies that have been pointed out, such as the current mismatch between supernovae and CMB determinations of the Hubble constant \cite{Bernal:2016gxb}, a slight hemispherical power asymmetry in the CMB \cite{PhysRevD.87.123005}, tension between low redshift weak lensing measurements from the CFHTLenS and CMB estimates of the current matter density, $\Omega_m$, and density fluctuations on 8 Mpc scales, $\sigma_8$, \cite{MacCrann:2014wfa}.  These examples are not meant to be exhaustive but instead to make the point that it is only by looking for deviations from the standard cosmological model that we increase our knowledge of, and focus our questions about, the physical nature of the universe.

In this work we investigate constraints to compensated isocurvature perturbations (CIPs) using observations of the CMB made by the \textit{Planck} satellite \cite{Adam:2015rua}.  The standard cosmological model predicts that the initial perturbations in the early universe are adiabatic (i.e., isentropic).  Several other types of initial perturbations may be established by non-standard processes (such as axion physics \cite{Linde:1996gt}, or alternatives to single-field slow-roll inflation such as the curvaton scenario \cite{Lyth:2002my}).  The most general set of these non-standard perturbations, called isocurvature perturbations, describe the `normal modes' of the early universe and as such evolve independently from each other and the standard adiabatic perturbations.  Most previous studies have placed constraints on the amplitude of pure isocurvature modes, finding that their amplitude cannot be larger than a few percent of that of the standard adiabatic perturbations \cite{Ade:2015lrj}.  On the other hand, CIPs are not pure modes, but are instead composed of a linear combination of baryon and cold dark matter (CDM) isocurvature.  The amplitude of these two modes are set so as to leave the total matter perturbation unchanged. CIPs are only weakly constrained by current data since the effects of CIPs on scales $k \lesssim 200\ {\rm Mpc}^{-1}$ appear at second order in the CIP amplitude \cite{Gordon:2009wx,Grin:2011nk,Grin:2011tf}. This makes the search for CIPs in current data sets, which probe scales $k  \lesssim 10\ {\rm Mpc}^{-1}$, particulalry challenging.
Moreover, CIPs are a natural prediction of certain non-standard inflationary theories, such as the curvaton model \cite{Lyth:2002my,He:2015msa,Smith:2015bln},
so searching for CIPs helps to shed light onto the physics of the early universe. 

There are several previous studies which have placed constraints on CIPs.  A CIP leads to fluctuations of the baryon to dark matter ratio which may be observed in the baryon fraction of galaxies \cite{Holder:2009gd} or in the detailed structure of the baryon acoustic oscillations \cite{Soumagnac:2016bjk}. It also generically leads to fluctuations in the primordial light element abundances \cite{Holder:2009gd}. These effects can be used to place constraints on CIPs for a variety of length-scales, from $\sim$ 1-100 Mpc.   

In the CMB, CIPs cause a modulation of the photon-baryon sound speed leading to second-order effects in both the CMB power spectrum and trispectrum.
Given that both CIPs and the weak lensing of the CMB are at second order \cite{Lewis:2006fu} it follows that they share mathematical similarities in how they affect the predicted structure of the CMB.  In particular, both cause smoothing of the small-scale CMB power spectrum \cite{Munoz:2016,Valiviita:2017fbx}, and generate a (non-Gaussian) connected CMB trispectrum \cite{Grin:2011nk,Grin:2011tf}. 

 An optimal estimator of CIPs using the CMB trispectrum was derived in Refs.~\cite{Grin:2011nk,Grin:2011tf,He:2015msa}. To evaluate this estimator on the \textit{Planck} data would be computationally intensive, requiring careful treatment of sky cuts and many simulations of mock CMB maps to obtain the relevant covariance matrices.  Instead, in this work we note that the standard estimator for the lensing potential power spectrum \cite{Okamoto:2003zw}, already implemented in the publicly data products from \textit{Planck} \cite{Ade:2015zua}, has sensitivity to the CIP field on large angular scales. We thus use current estimates of the lensing potential power spectrum from \textit{Planck} to search for CIPs and obtain
 limits to a scale-invariant CIP spectrum that are a factor of $\sim 3$ better than the limit from the CMB power spectrum alone: the root-mean-square CIP amplitude on CMB scales is $\Delta_{\rm rms}(R_{\rm CMB})^{2}\lesssim 4.3 \times 10^{-3}$ at 95\% confidence level (CL) . Using standard forecasting techniques, we find that we have extracted nearly all the information on CIPs that can be extracted from \textit{Planck} maps. For a future nearly-cosmic-variance-limited experiment like the CMB Stage-4, we find that an optimal analysis of the full trispectrum only improves on the lensing potential + CMB power spectrum analysis by a factor of $\sim 4$, driven mainly by polarization measurements.

Throughout this paper we use a fiducial cosmology that is spatially flat with parameters: $\Omega_b h^2 = 0.0222$, $\Omega_c h^2 = 0.1203$, $\Omega_\nu h^2 = 0.00064$ (corresponding to two massless neutrinos and one massive neutrino with $m = 0.06$ eV), $H_0 = 67.12\ {\rm km\ s^{-1}\ Mpc^{-1}}$, $A_s = 2.09 \times 10^{-9}$, $n_s = 0.96$, and $\tau = 0.065$. We begin with a summary of the physics of compensated isocurvature perturbations in Sec.~\ref{sec:cipsum}, and then review the effects of CIPs on the observed CMB fluctuations in Sec.~\ref{sec:cip_cmb}. Previous constraints to CIPs are summarized and explained in Sec.~\ref{sec:oldconstraints}. Our new CIP constraints, using a combination of the observed \textit{Planck} lensing-potential and CMB primary spectra, are presented in Sec.~\ref{sec:constraints}. We discuss the promise of more optimal estimators and future experiments in Sec.~\ref{sec:forecasts}, and conclude in Sec.~\ref{sec:conclusions}.

\section{Compensated isocurvature perturbations}
\label{sec:cipsum}

Solutions to the linearized Einstein equations, which describe perturbations to an otherwise isotropic and homogeneous universe, can be divided up into a set of normal modes each of which evolve independently (see, e.g., Ref.~\cite{Bucher:1999re}).  For example, a given Fourier mode of the density contrast ($\delta_i \equiv \delta \rho_i/\bar \rho_i$) of a species $i$ can be written as:
\begin{eqnarray}
    \delta_i(\vec k, \tau) &=& \sum_n \zeta_n(\vec k) A_n T^n_i(k, \tau),\\
    &=& \delta_i^{\rm ad} + \delta_i^{\rm b,iso} + \delta_i^{\rm c,iso}+ \cdots,
\end{eqnarray}
where $n$ denotes the type of initial conditions (i.e., adiabatic would be $n=0$, baryon isocurvature $n=1$ and so forth), $A_n$ gives the relative amplitude of the modes, $\zeta_n(\vec k)$ is a stochastic amplitude with zero mean, and $T^n_i(k, \tau)$ is the transfer function for each species and each type of initial condition. Similar equations can be written for the other perturbed moments of each species' stress-energy tensor.  The linearized Einstein equations determine the behavior of the transfer functions $T^n_i(k, \tau)$. 

For any set of initial conditions the relative entropy perturbation between any species and photons is given by 
\begin{equation}
S_{i} \equiv 3 (\xi_i - \xi_\gamma) = - 3 \mathcal{H} \left(\frac{\delta \rho_i}{\rho'_i}- \frac{\delta \rho_\gamma}{\rho_{\gamma}'}\right), 
\end{equation}
where $\xi_i$ is the curvature perturbation in a gauge where that species is uniform, $\mathcal{H} \equiv a'/a$, and the prime indicates a derivative with respect to conformal time, $\tau$.  
We can simplify this expression by noting that the continuity equation applied to the background density is 
\begin{equation}
\rho_i' = - 3 \mathcal{H} \rho_i (1+w_i), 
\end{equation}
so that 
\begin{equation}
S_{i} = \frac{1}{1+w_i} \delta_i - \frac{3}{4} \delta_\gamma.
\end{equation}
We consider a universe filled with photons ($\gamma$), neutrinos ($\nu$), baryons ($b$), and cold dark matter (CDM, $c$) (we also consider a cosmological constant, $\Lambda$, which does not cluster).  The \emph{total} matter ($m$) is the sum of the baryons and CDM.  The standard adiabatic initial conditions have $S_i = 0$ for all species; initial conditions which have $S_{i} \neq 0$ for some species as well as leave the Ricci scalar curvature of the universe unperturbed are called isocurvature perturbations and are independent of the adiabatic mode \cite{Bond:1984fp,Kodama:1986ud,Bucher:2000kb}.

A compensated isocurvature perturbation (CIP) is a linear combination of baryon and CDM isocurvature which has $S_\nu=S_m = 0$.  The evolution of the perturbations is then determined by the initial values of the linear combinations \cite{Bucher:2000kb}
\begin{eqnarray}
\delta_\gamma &=& \delta_\gamma^{\rm ad}, \\
\delta_\nu &=& \delta_\nu^{\rm ad},\\
\delta_c &=&  \delta_c^{\rm ad}+ \delta_c^{\rm CIP} ,\\
\delta_b &=& \delta_b^{\rm ad} + \delta_b^{\rm CIP},\\
\delta_m &=& R_b \delta_b+ R_c \delta_c  = \delta_m^{\rm ad},
\end{eqnarray}
where 
\begin{eqnarray}
R_c &\equiv& \frac{\rho_c}{\rho_c + \rho_b}, \\
R_b &\equiv& \frac{\rho_b}{\rho_c + \rho_b}.
\end{eqnarray}

Although the radiation and total matter perturbations are initially purely adiabatic, isocurvature fluctuations can be generated by pressure gradients in the baryon fluid. Additionally, until these pressure gradients are significant the CIP modes do not evolve in time. As pointed out in Ref.~\cite{Gordon:2009wx}, this pressure support is only significant wave numbers beyond the baryon sound horizon: $k \gtrsim 200\ {\rm Mpc}^{-1}$. Thus on larger scales, CIP modes do not evolve in time and can be treated as a function of position only. On scales larger than the baryon sound horizon we define the stochastic CIP field as $\Delta (\vec k) \equiv \delta_b^{\rm CIP}(\vec k)$ so that $\delta_c^{\rm CIP}(\vec k) = -\Omega_b/\Omega_c \Delta(\vec k)$. 
Since the CIP modes do not evolve in time, we can write 
\begin{equation}
\Delta_{\rm CIP}(\vec k) =  \zeta_{\rm CIP}(\vec k),
\end{equation}
so that in the presence of a CIP the baryon and CDM densities are modulated by $\Delta$:
\begin{eqnarray}
\rho_b(\vec x) = \bar{\rho}_b [1+\Delta(\vec x)],\\
\rho_c(\vec x) = \bar{\rho}_c\left[1-\frac{\bar{\rho}_b}{\bar{\rho}_c} \Delta(\vec x)\right],
\end{eqnarray}
where $\bar{\rho}_b/\bar{\rho}_c = \bar{\Omega}_b/\bar{\Omega}_c \simeq 0.2$ is the unperturbed (homogeneous) ratio of the baryon to cold dark matter density.  

The CIP field can have an arbitrary correlation with the primordial curvature perturbation \cite{Gordon:2009wx,He:2015msa}.  Most of the previous work on CIPs has assumed $\Delta$ is uncorrelated with the primordial curvature perturbation (e.g., Refs.~\cite{Grin:2011nk,Grin:2011tf,Grin:2013uya,Munoz:2016}).  This assumption greatly simplifies the effects of a CIP since in this case only auto-correlations of $\Delta$ are non-zero. On the other hand fully correlated CIPs are a natural prediction of the curvaton scenario \cite{Smith:2015bln, He:2015msa}.  The additional correlations present in this case lead to a greater sensitivity to the CIP field \cite{He:2015msa,Abazajian:2016yjj}, and in future work we will leverage this sensitivity to obtain forecasts and constraints to curvaton-inspired CIPs. For the rest of this paper, we consider only  CIP which are uncorrelated with the primordial curvature perturbation.

\section{CIPs and the CMB}
\label{sec:cip_cmb}

The main effect of CIPs on the CMB is the spatial modulation in the photon/baryon sound speed \cite{He:2015msa}.  In particular, in the presence of a CIP the acoustic waves that generate the structure of the observed CMB anisotropies propagate through an inhomogeneous medium with a sound speed that varies as 
\begin{equation}
c_s^2(\vec x) = \frac{1}{3} \left(1+\frac{3}{4}\frac{\bar{\rho}_b [1+\Delta(\vec x)]}{\bar{\rho}_\gamma}\right)^{-1}
\simeq (\bar{c}_s)^2 \left[1-\frac{3 \bar{\rho}_b}{3 \bar{\rho}_b + 4 \bar{\rho}_\gamma} \Delta(\vec x)\right].
\end{equation}
Additionally, the modulation of $\rho_b$ leads to a spatial variation in the visibility function at decoupling.  For CIP scales smaller than the acoustic horizon at decoupling ($L \gtrsim 100$), the effects of the CIP modulation are suppressed \cite{He:2015msa}, and so we only include CIP multipoles at scales larger than the acoustic horizon. We can write the angular CIP field 
\begin{equation}
\Delta(\hat n) = \Delta(\chi_* \hat n), 
\end{equation}
where $\chi_*$ is the comoving distance to the surface of last scattering.  Writing the Fourier transform of the CIP field as 
\begin{equation}
\Delta(\vec x) = \int \frac{d^3 k}{(2\pi)^{3}} \Delta(\vec k) e^{i \vec k \cdot \vec x},
\end{equation}
its power spectrum is given by
\begin{equation}
\langle \Delta(\vec k) \Delta^*(\vec k\ ')\rangle = (2\pi)^3 \delta_D^{(3)}(\vec k - \vec k\ ')P_{\Delta \Delta}(k),
\end{equation}
and we can compute the root-mean square CIP amplitude over some length-scale $R$ as
\begin{equation}
\Delta^2_{\rm rms}(R) = \frac{1}{2\pi^2} \int k^2 dk[3 j_1(kR)/(kR)]^2 P_{\Delta \Delta}(k).
\label{eq:DeltaA}
\end{equation}
Finally, the CIP angular distribution at the last-scattering surface will be given by 
\begin{equation}
\Delta_{L M} =  \frac{4 \pi i^L}{(2\pi)^{3/2}}\int d^3 k \Delta(\vec k) j_L(k \chi_*) Y^*_{L M}(\hat k),
\end{equation}
which gives rise to an angular power spectrum
\begin{equation}
C_L^{\Delta \Delta} = \langle \Delta_{L M} \Delta^*_{L M}\rangle =  \frac{2}{\pi}
\int k^2d k P_{\Delta \Delta}(k) j^2_L(k \chi_*) = \frac{A_{\rm CIP}}{\pi L(L+1)},
\end{equation}
and we assume that the CIP power spectrum is scale invariant: $P_{\Delta \Delta}(k) = A_{\rm CIP}/k^3$. 
These approximations allow us to write $\Delta_{\rm CIP}(\hat n)$ with \begin{equation}
\Delta^2_{\rm rms}(R_{\rm CMB}) \equiv \langle \Delta^{2}(\hat n) \rangle  = \sum_{L=1}^{100} \frac{2L+1}{4\pi} C_L^{\Delta \Delta} \simeq \frac{A_{\rm CIP}}{4},\label{eq:DeltatoACIP}
\end{equation}
where we have truncated the sum at $L =100$ since, as stated before, the CIP modulation damps away on scales smaller than the acoustic horizon at decoupling. Finally, we find the expression 
\begin{equation}
C_L^{\Delta \Delta} \simeq \frac{4}{\pi}\frac{\Delta_{\rm rms}^2(R_{\rm CMB})}{L(L+1)}.
\end{equation}

The effects of a CIP modulation on the anisotropies of the CMB are most clearly understood using a flat-sky approximation. For the discussion here we only present results for the temperature anisotropies. When searching for CIPs, we use both temperature and polarization with the full-sky expressions found in Appendix \ref{append:fullsky}. 

Weak gravitational lensing and CIPs can be thought of as a modulation of a `background' CMB anisotropy $T(\hat n)$ giving an observed anisotropy $T_{\rm obs}(\hat n)$.  
In the presence of both weak gravitational lensing and CIPs the temperature anisotropies are given by 
\begin{eqnarray}
T_{\rm obs}(\hat n) &=&T\left[\hat n + \vec{\nabla} \phi(\hat n), \Delta(\hat n)\right],\\
&\simeq& \tilde T(\hat n) + \nabla_i \phi \nabla^i  T + \Delta(\hat n) \frac{\partial  T}{\partial \Delta}\bigg|_{\Delta = 0} + \frac{1}{2}\left( \nabla_i \phi \nabla_j \phi \nabla^i \nabla^j  T + \Delta^{2}(\hat n) \frac{\partial^2  T}{\partial \Delta^2}\bigg|_{\Delta = 0}\right) + \cdots,
\end{eqnarray} where the terms proportional to derivatives of $\phi(\hat{n})$ are standard lensing contributions as first derived in Refs.~\cite{Zaldarriaga:1998te,Hu:2000ee} and described in Ref.~\cite{lewis2006weak}.
Additionally, one must include a noise term, so then the total observed temperature at each point on the sky can be written $T^t(\hat n) = T(\hat n) + T^N(\hat n)$, where we assume that we are using beam-deconvolved maps. This leads to an estimated power spectrum for the beam-deconvolved map 
\begin{equation}
C_{l}^{TT,{\rm t}} = C_{l}^{TT} + C_{l}^{TT,N},
\label{eq:obsmaps}
\end{equation}
where the inverse-variance-weighted sum over all channels $i$ gives
\begin{equation}
C_{l}^{TT,N} = \left(\sum_i w^{-2}_{T,i} e^{-l^2 \sigma_{b,i}^2}\right)^{-1},\label{eq:knoxknoise}
\end{equation}
with $\sigma_{b,i} \equiv \theta_i/\sqrt{8\ln 2}$, $\theta_i$ is the full-width half-maximum, and $w_{T,i}$ is the weight per solid angle for each channel. 

Taking the Fourier transform of the temperature map we have 
\begin{eqnarray}
T_{\rm obs}(\vec l) \equiv \int d^2 \hat n T_{\rm obs}(\hat n) e^{- i \vec l \cdot \hat n} 
= T(\vec l) + \delta T^{(1)}(\vec l) + \delta T^{(2)}(\vec l),
\end{eqnarray} 
where the first- and second-order terms are given by
\begin{eqnarray}
\delta T^{(1)}(\vec l) &=& \int \frac{d^2 l\ '}{(2\pi)^2} T(\vec l\ ') \phi(\vec l- \vec l\ ') L_{\phi}^{(1)}(\vec l,\vec l\ ') +\frac{\partial \tilde T(\vec l\ ')}{\partial \Delta}\bigg|_{\Delta = 0}\Delta(\vec l-\vec l\ '),\\
\delta T^{(2)}(\vec l) &=& \frac{1}{2}\int \frac{d^2 l\ '}{(2\pi)^2}\frac{d^2 l\ ''}{(2\pi)^2} T(\vec l\ ') \phi(\vec l\ '') \phi(\vec l-\vec l\ '-\vec l\ '') L_{\phi}^{(2)}(\vec l, \vec l\ ',\vec l\ '') +\frac{\partial^2 \tilde T(\vec l\ ')}{\partial \Delta^2}\bigg|_{\Delta = 0} \Delta(\vec l\ '')\Delta(\vec l-\vec l\ '-\vec l\ ''),
\end{eqnarray}
and for simpler notation we define
\begin{eqnarray}
L^{(1)}_{\phi}(\vec l,\vec l\ ') &\equiv&-[(\vec l-\vec l\ ')\cdot \vec l\ '],\\
L^{(2)}_{\phi}(\vec l, \vec l',\vec l\ '') &\equiv& - \left[\vec l\ ''\cdot \vec l\ '\right] \left[(\vec l\ ''+ \vec l\ ' - \vec l)\cdot \vec l\ '\right].
\end{eqnarray}

The observed power spectrum is found by
\begin{equation}
\langle T_{\rm obs}(\vec l) T_{\rm obs}^*(\vec l\ ')\rangle\equiv (2\pi)^2\delta_D^{(2)}(\vec l +\vec l ')C^{TT,{\rm obs}}_l.
\end{equation}  
From these expressions it is straightforward to show that in the presence of both lensing and a CIP the observed power spectrum becomes 
\begin{eqnarray}
C^{TT, {\rm obs}}_l &=& C^{TT}_l\left[1 -\int \frac{d^2 L}{(2\pi)^2} C_L^{\phi \phi} (\vec L \cdot \vec l)^2\right]+\int \frac{d^2 L}{(2\pi)^2} C_{|\vec l - \vec L|}^{TT}C_L^{\phi \phi}[(\vec l - L)\cdot \vec L]^2\\ \nonumber &+& \int \frac{d^2L}{(2\pi)^2}  C^{dT,dT}_{| \vec l - \vec L|} C^{\Delta \Delta}_L 
+ C^{T,d^2 T}_l \int \frac{d^2L}{(2\pi)^2} C_L^{\Delta \Delta},\\
&=&C^{TT}_l+\delta C^{TT,\phi}_l + \delta C^{TT,\Delta}_l,
\end{eqnarray} 
where we have defined 
\begin{eqnarray}
\VEV{\phi(\vec l) \phi(\vec l')} &\equiv& (2\pi)^2 C_l^{\phi \phi} \delta_D^{(2)}(\vec  l - \vec l'),\\
\bigg\langle \frac{\partial \tilde T(\vec l)}{\partial \Delta} \frac{\partial \tilde T(\vec l')}{\partial \Delta}\bigg\rangle &\equiv&(2\pi)^2 C^{dT,dT}_{l}\delta_D^{(2)}(\vec l +\vec l '),\\
\bigg\langle \tilde T(\vec l) \frac{\partial^2 \tilde T(\vec l')}{\partial \Delta^2}\bigg\rangle &\equiv&(2\pi)^2 C^{T,d^2T}_{l}\delta_D^{(2)}(\vec l +\vec l ').
\end{eqnarray}
Since the CIP field modulates the baryon density it follows that $C^{T,d^2T}_{l}$ is only significant on scales smaller than the acoustic horizon, $l \gtrsim 100$. Furthermore, for a scale-invariant power spectrum $C_L^{\Delta \Delta} \propto 1/L^2$, which peaks at small $L$. This separation in scales allows us to write the CIP contribution to the observed CMB power spectrum as 
\begin{equation}
\delta C^{TT,\Delta}_l=\int \frac{d^2L}{(2\pi)^2}  C^{dT,dT}_{| \vec l - \vec L|} C^{\Delta \Delta}_L 
+ C^{T,d^2 T}_l \int \frac{d^2L}{(2\pi)^2} C_L^{\Delta \Delta} \simeq \frac{1}{2} \Delta_{\rm rms}^2(R_{\rm CMB}) \frac{\partial^2 C_l^{TT}}{\partial \Delta^2},\label{eq:flatClobs}
\end{equation}
where in the flat-sky approximation $\Delta_{\rm rms}^2(R_{\rm CMB}) \equiv \int_1^{100} d^2L/(2\pi) C_L^{\Delta \Delta}$. This result also holds for the full-sky expressions, as discussed in Appendix \ref{append:fullsky}. 

\begin{figure}[!ht]
\resizebox{!}{12cm}{\includegraphics{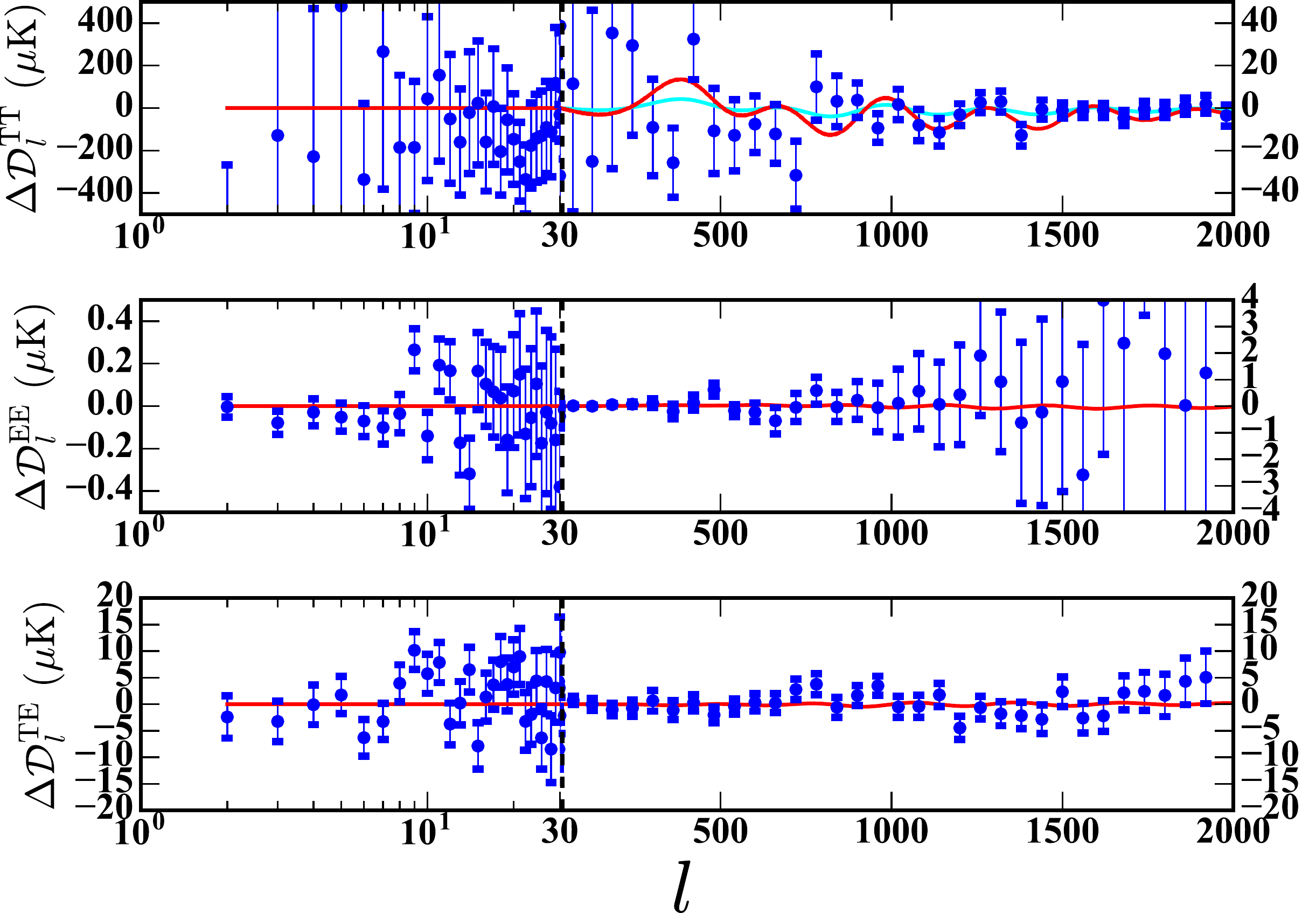}}
\caption{A comparison between the difference between a the standard $\Lambda$CDM CMB power spectra and a one which has been modulated by a CIP mode.  Each panel shows the binned residuals $\Delta \mathcal{D}^{XX'}_{l} \equiv l(l+1) \Delta C^{XX'}_{l}/(2\pi)$ from the Planck satellite (see Ref.~\cite{Aghanim:2015xee} for details on the binning procedure).  Two CIP mode amplitudes are shown: the red curve shows the CIP mode which saturates the 95\% CL upper limit using both the Planck measurements of the temperature and polarization CMB power spectra, $\Delta_{\rm rms}^2(R_{\rm CMB}) = 1.39 \times 10^{-2}$; the cyan curve (only visible in the top panel) shows the CIP mode which saturates the 95\% CL upper limit using both the Planck measurements of the CMB power spectra as well as the lensing potential power spectrum, $\Delta_{\rm rms}^2(R_{\rm CMB}) = 4.3 \times 10^{-3}$. Note that the horizontal scale is logarithmic up to $l = 29$ and then is linear; the vertical scale on the left and right-hand sides are different.}
\label{fig:Planck_diff}
\end{figure}

Both weak lensing and the CIP modulation cause a smoothing of the CMB power spectra on scales smaller than the acoustic horizon \cite{Munoz:2016}.  We show the residual of the fiducial $\Lambda$CDM power spectrum with the CIP-modulated CMB power spectra, $\Delta \mathcal{D}_l^{XX'} \equiv l(l+1) \Delta C_l^{XX'}/(2\pi)$ with $X=\{T,E\}$, along with the Planck measurements in Fig.~\ref{fig:Planck_diff}.  This figure makes it clear how the measurements of the CMB power spectrum are sensitive to the presence of a CIP mode. In the residuals the additional smoothing of the peaks leads to an oscillatory structure, which is most apparent in the temperature power spectrum. 

The CIP modulation also produces a contribution to correlations beyond the CMB power spectrum.  In particular, Refs.~\cite{Grin:2011nk,Grin:2011tf,Grin:2013uya,He:2015msa} construct an optimal estimator for $\Delta_{LM}$ from the connected part of the CMB four-point correlation, the trispectrum.  The analysis of the CMB trispectrum is far from trivial, so here we utilize the fact that estimates of the lensing potential power spectrum, $\phi$, are also built out of the connected part of the CMB trispectrum \cite{Okamoto:2003zw,2011PhRvL.107b1301D,Ade:2015zua}.  In the presence of CIPs, the estimator used to reconstruct the lensing potential power spectrum gains an additional contribution proportional to $\Delta_{\rm rms}^2(R_{\rm CMB})$.  

Both weak gravitational lensing and CIPs introduce a non-Gaussian connected trispectrum
\begin{equation}
\langle T(\vec l_1) T(\vec l_2) T(\vec l_3) T(\vec l_4)\rangle_c =  \mathcal{T}(\vec l_1,\vec l_1,\vec l_1,\vec l_1) \delta^{(2)}(\vec l_1+\vec l_2+ \vec l_3 + \vec l_4).
\end{equation}
The dominant contributions to the connected trispectrum are given by \cite{PhysRevD.67.123507}
\begin{eqnarray}
  \mathcal{T}(\vec l_1,\vec l_2,\vec l_3,\vec l_4) &\simeq& 
    C_{|\vec l_1+\vec l_2|}^{\phi \phi} f_{TT}(\vec l_1,\vec l_2)f_{TT}(\vec l_3, \vec l_4) +C_{|\vec l_1+\vec l_2|}^{\Delta \Delta} h_{TT}(\vec l_1,\vec l_2)h_{TT}(\vec l_3, \vec l_4)\nonumber \\
    &+& C_{|\vec l_1+\vec l_3|}^{\phi \phi} f_{TT}(\vec l_1,\vec l_3)f_{TT}(\vec l_2, \vec l_4) +C_{|\vec l_1+\vec l_3|}^{\Delta \Delta} h_{TT}(\vec l_1,\vec l_3)h_{TT}(\vec l_2, \vec l_4)\nonumber \\
    &+&C_{|\vec l_1+\vec l_4|}^{\phi \phi} f_{TT}(\vec l_1,\vec l_4)f_{TT}(\vec l_2, \vec l_3) +C_{|\vec l_1+\vec l_4|}^{\Delta \Delta} h_{TT}(\vec l_1,\vec l_4)h_{TT}(\vec l_2, \vec l_3),
    \label{eq:4point}
\end{eqnarray}
where 
\begin{eqnarray}
 f_{TT}(\vec l_1, \vec l_2) &\equiv&  [(\vec l_1 + \vec l_2)\cdot \vec l_1] C_{l_1}^{TT} + [(\vec l_1 + \vec l_2) \cdot \vec l_2]C_{l_2}^{TT},\\
  h_{TT}(\vec l_1, \vec l_2) &\equiv&C^{T,dT}_{l_1} + C^{T,dT}_{l_2}.\label{eq:hTT}
\end{eqnarray}
We can construct an estimator for the lensing potential power spectrum, $C_L^{\phi \phi}$, out of the 4-point correlation function of the temperature, E- and B-mode polarization of the CMB. A nearly optimal, inverse variance-weighted, estimator has been derived \cite{Okamoto:2003zw} and applied to the Planck CMB maps \cite{Ade:2015zua}. 
As shown in Eq.~(\ref{eq:4point}), in the presence of a CIP modulation this estimator includes contributions from the CIP field. 

In the flat-sky approximation, a minimum-variance estimator for the weak-lensing deflection field, $d(\vec L) \equiv \sqrt{L(L+1)}\phi(\vec L)$ can be built out of off-diagonal correlations of the CMB temperature fluctuations, and is \cite{Hu:2001kj,Okamoto:2003zw,Ade:2015zua} 
\begin{equation}
\hat{d}_{TT}(\vec L) \equiv \frac{i \vec L A_{TT}(L)}{L^2} \int \frac{d^2 l_1}{(2\pi)^2} T(\vec l_1) T(\vec l_2) F_{TT}(\vec l_1,  \vec l_2),
\label{eq:dhat}
\end{equation}
where 
\begin{eqnarray}
\vec l_2 &\equiv& \vec L - \vec l_1,\\
F_{TT}(\vec l_1,\vec l_2) &\equiv& \frac{f_{TT}(\vec l_1, \vec l_2)}{2 C_{l_1}^{TT,{\rm t}} C_{l_2}^{TT,{\rm t}}},\\
A_{TT}(L) &\equiv& L^2 \left[\int \frac{d^2 l_1}{(2\pi)^2} f_{TT}(\vec l_1, \vec l_2) F_{TT}(\vec l_1,  \vec l_2)\right]^{-1}.
\end{eqnarray}
The lensing potential power spectrum is estimated from the expectation value of the estimator in Eq.~\eqref{eq:dhat}
\begin{equation}
\hat{C}_L^{\phi \phi} \equiv \frac{1}{2L+1}\sum_{M=-L}^{L} \frac{\hat{d}^2_{TT}(\vec L)}{L(L+1)} - B(L),
\end{equation}
where $\pi M/L$ is the angular coordinate of $\vec L$ and $B(L)$ are the standard Gaussian and non-Gaussian contributions to the full four-point correlation \cite{PhysRevD.67.123507,2011PhRvL.107b1301D,Ade:2015zua}.

\begin{figure}[!ht]
\resizebox{!}{7cm}
{\includegraphics{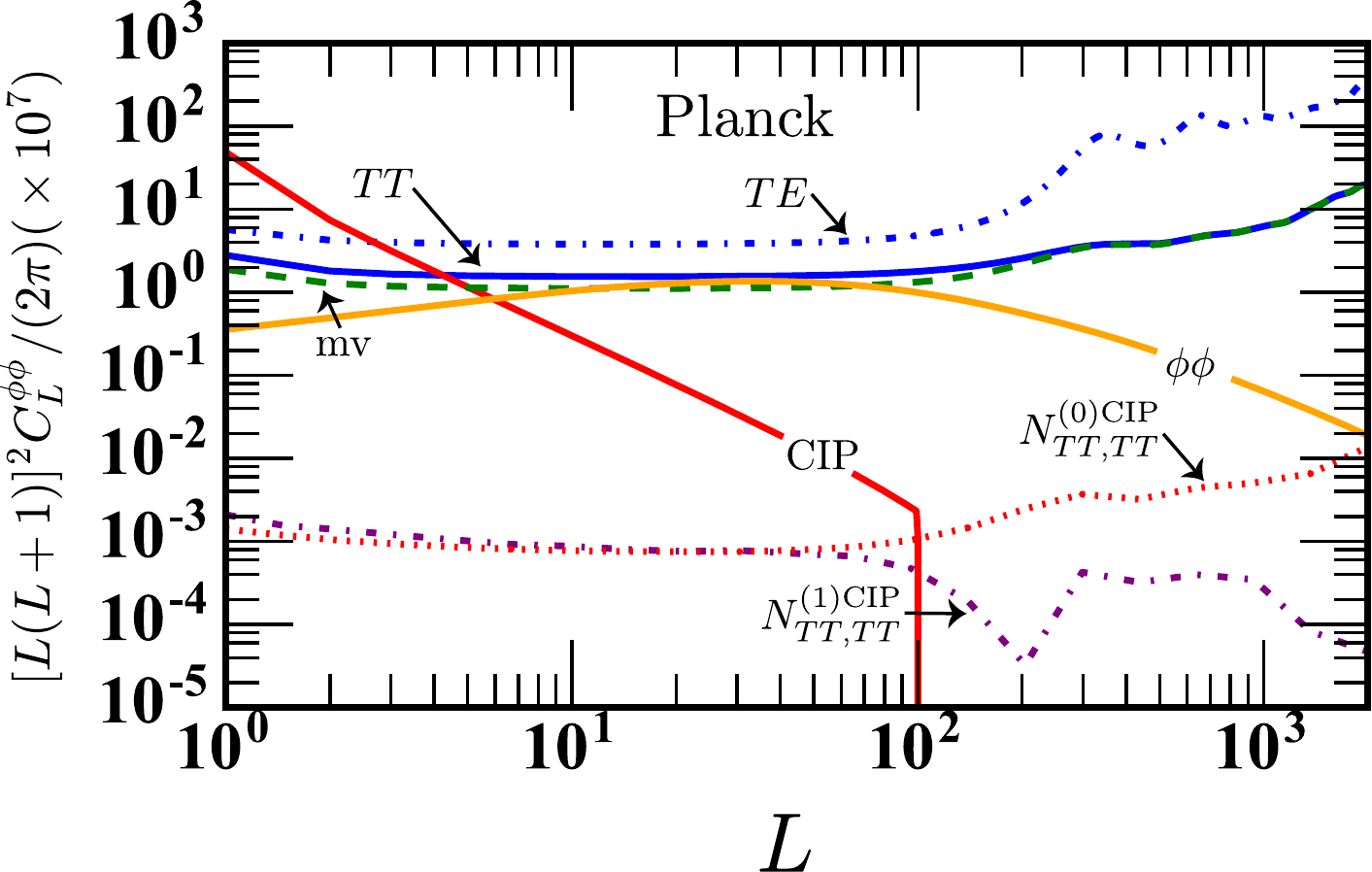}}
\caption{The CIP contribution to the expectation value of the standard lensing potential power spectrum estimator for Planck with $\Delta_{\rm rms}^2 = 4.3 \times 10^{-3}$. The solid orange curve shows the lensing potential power spectrum. The blue curves show the Gaussian noise from the $TT$ estimator (solid) and the $TE$ estimator (dot-dashed). The dashed green curve shows the noise of the minimum variance estimator. The dashed red curve shows the residual CIP contribution to the Gaussian noise if it is not subtracted from the signal and the purple dot-dashed curve shows the non-Gaussian (i.e., connected part of the trispectrum) CIP contribution. The solid red curve shows the dominant CIP contribution to the estimator.}
\label{fig:clpp_contrib_Planck}
\end{figure}

In the presence of a CIP modulation, the standard lensing potential power-spectrum estimator has expectation value
\begin{eqnarray}
L^2\langle  \hat{C}^{\phi \phi}_L\rangle  &=& L^2 C^{\phi \phi}_{L}+L^2C^{\Delta \Delta}_{L} [Q^{TT}(L)]^2+ N^{(0){\rm CIP}}_{TT,TT}(L)+N^{(1){\rm CIP}}_{TT,TT}(L),\label{eq:expect}
\end{eqnarray}
where the last two terms are the CIP-induced Gaussian and non-Gaussian bias, and 
\begin{equation}
Q^{TT}(L) \equiv \frac{\int\frac{d^2 l_1}{(2\pi)^2}h_{TT}(\vec l_1, \vec l_2)F_{TT}(\vec l_1,\vec l_2) }{\int \frac{d^2 l_1}{(2\pi)^2} f_{TT}(\vec l_1,\vec l_2) F_{TT}(\vec l_1,\vec l_2)}
\label{eq:QTT}
\end{equation}
The biases can be written to leading order in the CIP amplitude as \cite{PhysRevD.67.123507}
\begin{eqnarray}
 N^{(0){\rm CIP}}_{TT,TT}(L) &=& \frac{1}{2} \Delta_{\rm rms}^2(R_{\rm CMB}) \left(\frac{A(L)}{L}\right)^2 \int \frac{d^2 l_1}{(2\pi)^2} f_{TT}(\vec l_1, \vec l_2) F_{TT}(\vec l_1,\vec l_2)  \left[\frac{1}{{C}_{l_1}^{TT}} \frac{\partial C^{TT}_{l_1}}{\partial \Delta^2} + \frac{1}{{C}_{l_2}^{TT}} \frac{\partial C^{TT}_{l_2}}{\partial \Delta^2}\right],\\
 N^{(1){\rm CIP}}_{TT,TT}(L) &=&\frac{A^2_{TT}(L)}{L^2} \int \frac{d^2 l_1}{(2\pi)^2} \frac{d^2 l_1'}{(2\pi)^2}F_{TT}(\vec l_1,\vec l_2)F_{TT}(\vec l_1',\vec l_2')\bigg[ C^{\Delta \Delta}_{|\vec l_1-\vec l_1'|} h_{TT}(-\vec l_1,\vec l_1') h_{TT}(-\vec l_2,\vec l_2')\nonumber \\
&+& C^{\Delta \Delta}_{|\vec l_1-\vec l_2'|} h_{TT}(-\vec l_1,\vec l_2') h_{TT}(-\vec l_2,\vec l_1')\bigg].
\end{eqnarray}
In practice, the Planck lensing analysis uses a combination of observed and simulated CMB maps to subtract off the Gaussian bias \cite{Ade:2015zua}.  Since the simulated maps do not include a CIP contribution, some fraction of the CIP contribution to the Gaussian bias may not be fully subtracted. On the other hand, the CIP non-Gaussian bias is not subtracted off in the standard analysis. See Appendix \ref{sec:CIPeffects} for a more complete derivation of this expectation value in the presence of a CIP modulation. 

The full analysis \cite{Ade:2015zua}, which includes both CMB temperature and polarization maps, computes a minimum variance (mv) estimator from all possible CMB map auto- and cross-correlations, as discussed in detail in Appendix \ref{sec:CIPeffects} and \ref{append:fullsky}.

In the case of Planck, the lensing estimator is made using maps constructed from the 143 GHz and 217 GHz channels. 
\begin{table}[!ht]
\begin{center}
\noindent\adjustbox{max width=1.2\textwidth}{%
\begin{tabular}{p{15mm}ccc}
\hline\hline
Channel & $\theta$ (arcmin) & $w_T\ (\mu{\rm K}\ {\rm arcmin})$ & $w_P\ (\mu{\rm K}\ {\rm arcmin})$ \\\noalign{\smallskip}\hline\noalign{\smallskip}
143 GHz & 7 & 30 & 60  \\\noalign{\smallskip}
217 GHz & 5 & 40 & 95\\\noalign{\smallskip}
CMB-S4 & 3 & 1 & 1.4
\\\noalign{\smallskip}
\hline\hline
\end{tabular}}
\caption{Planck sensitivity in the 143 and 217 GHz channels to temperature and polarization at the two frequencies used to estimate the lensing potential \cite{Ade:2015zua,Adam:2015vua}. The last line gives the sensitivity for CMB-S4, a proposed next generation CMB telescope \cite{Abazajian:2016yjj}.} 
\end{center}
\label{table:plancksens}
\end{table}
We furthermore note that the Planck analysis uses a bandpass filter in harmonic space to restrict the power spectrum multipoles to $100 \leqslant  l \leqslant 2048 $.

The CIP contribution to the lensing potential estimator for the Planck satellite is shown in the solid red curve in Fig.~\ref{fig:clpp_contrib_Planck} for $\Delta_{\rm rms}^2(R_{\rm CMB}) = 4.3 \times 10^{-3}$. We can see that both the Gaussian and non-Gaussian CIP contributions to the lensing estimator (dotted red and dot-dashed purple curves) are negligible compared to the lensing potential power spectrum (solid yellow); the only significant CIP contribution is given by $L^2C^{\Delta \Delta}_{L} [Q^{TT}(L)]^2$. In the dashed green curve we also show the Gaussian noise in a minimum-variance estimator using Planck.

Given our assumption of a scale-invariant CIP power spectrum we can write the CIP contribution to the lensing potential estimator as 
\begin{equation}
L^2C^{\Delta \Delta}_{L} [Q^{TT}(L)]^2 = \frac{A_{\rm CIP}}{\pi} [Q^{TT}(L)]^2,
\end{equation}
where in the flat-sky limit we have used the fact that $C^{\Delta \Delta}_{L} = A_{\rm CIP}/(\pi L^2)$. For small values of $L$ the function $h_{TT}(\vec l_1, \vec L - \vec l_1)$ [see Eq.~(\ref{eq:hTT})], which goes into computing the $Q_L$ above, is nearly independent of $L$. Using the roughly sinusoidal solutions for wave amplitudes in the baryon-photon plasma, we find that the leading order contribution (in $L/l_{\rm max}$) to $h_{TT}(\vec l_1, \vec L - \vec l_1)$ is
\begin{equation}
h_{TT}( \vec l_1, \vec L - \vec l_1) \simeq 2 C_{l_1}^{T,dT} \simeq 2 \frac{A_s}{l_1^2} \frac{\partial \ln c_s}{\partial \Delta} [1-\cos\left(2l_1/l_{\rm hor}\right)],
\end{equation}
where $l_{\rm hor} \simeq 100$ is the angular scale of the acoustic horizon at decoupling,
\begin{equation}
\frac{\partial \ln c_s}{\partial \Delta} = -\frac{3 \Omega_b}{6 \Omega_b + 8 (1+z_{\rm dec} )\Omega_{\gamma}}
\end{equation}
is the derivative of the photon-baryon sound speed at decoupling, and $A_s$ is the amplitude of the primordial scalar perturbations.  
\begin{figure}[!ht]
\begin{center}
\resizebox{!}{7cm}{\includegraphics{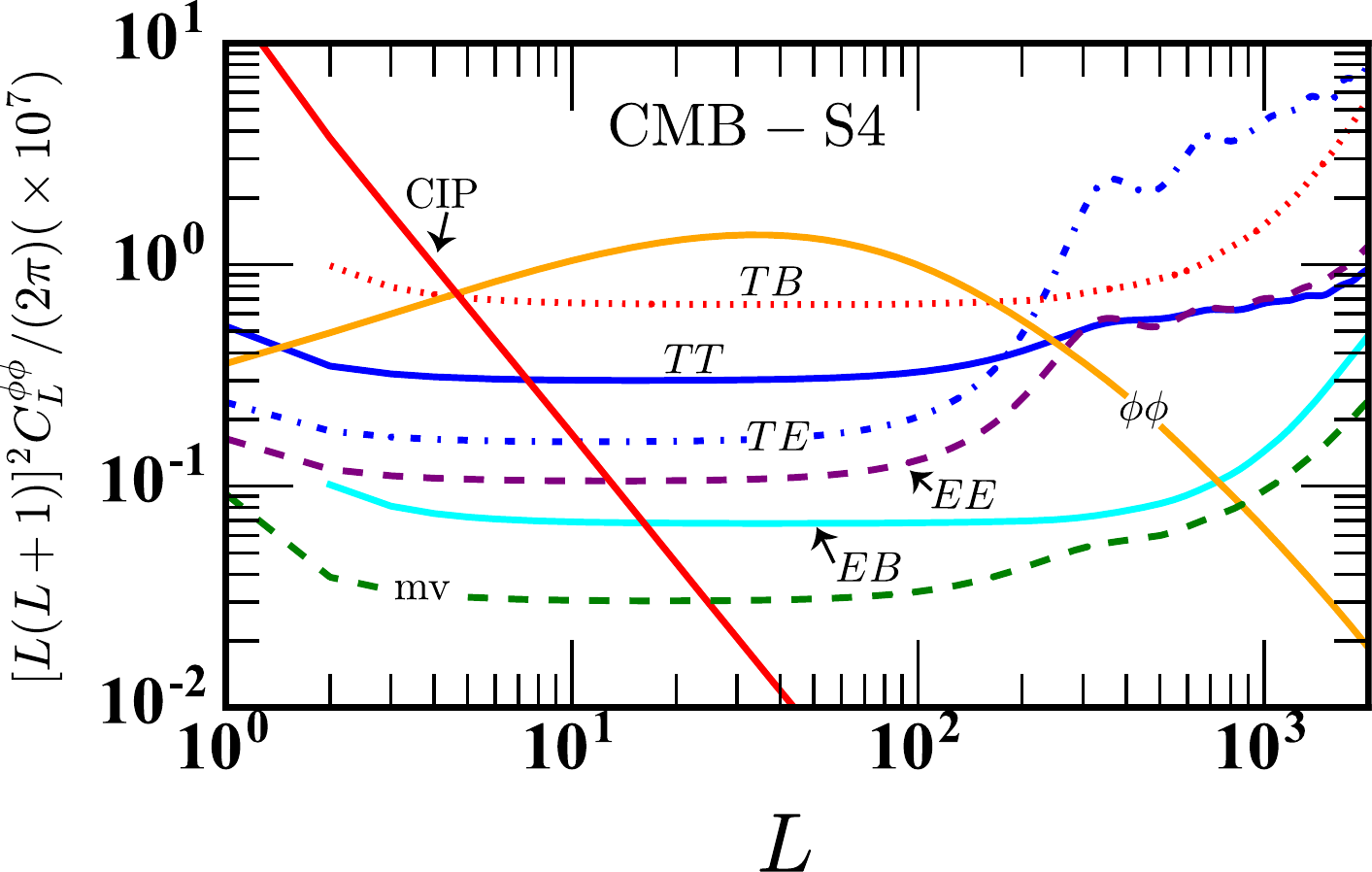}}
\caption{The CIP contribution to the expectation value of the standard lensing potential power spectrum estimator for a cosmic variance limited estimate of $C_L^{\phi \phi}$ with $\Delta_{\rm rms}^2(R_{\rm CMB}) = 4.3 \times 10^{-3}$. The expected $C_L^{\phi \phi}$ is shown in solid orange. The Gaussian noise associated with the $TT$, $TE$, $TB$, $EB$, and $EE$ estimators are shown in the solid blue, dashed-dot blue, dotted red, solid cyan, and dashed purple lines, respectively. The minimum variance estimator is the dotted green curve and the solid red curve shows the CIP contribution.}
\label{fig:clpp_contrib_CMBS4}
\end{center}
\end{figure}
Furthermore, the weight function $F_{TT} (\vec l_1, \vec L - \vec l_1) \propto 1/C_{l_1}^{TT}$, grows as $l_1^2$ so that the dominant terms which contribute to $Q^{TT}(L)$ have $L \ll l_1$ and we can write to leading order in $L/l_{\rm max}$ 
\begin{eqnarray}
f_{TT}(\vec l_1, \vec L- \vec l_1) &\simeq& L^2 C_{l_1}^{TT},\\
F_{TT} (\vec l_1, \vec L - \vec l_1) &\simeq& \frac{L^2}{2C^{TT}_{l_1}}.
\end{eqnarray}
Assuming that $l_{\rm max} \gg (l_{\rm hor},l_{\rm min})$  the CIP contribution is approximately given by
\begin{equation}
Q^{TT}(L) \simeq \frac{2}{L^2} \frac{\partial \ln c_s}{\partial \Delta}.
\label{eq:QTTapprox}
\end{equation}
Along with Eq.~(\ref{eq:expect}) this shows that the CIP contribution scales as $1/L^2$ relative to $L^4 C_L^{\phi \phi}$ and therefore dominates on large scales (i.e., small $L$). 

The approximate expression for the CIP contribution to the lensing potential power spectrum estimator in Eq.~(\ref{eq:QTTapprox}) allows us to estimate the extent to which this contribution varies with cosmological parameters. Given that $c_s$ depends only on the baryon density, we can see that variation of $\Omega_b$ causes the largest variation in the CIP contribution. In particular we expect a variation of $Q^{TT}(L)$ of order 
\begin{equation}
\frac{\delta Q^{TT}(L)}{Q^{TT}(L)} \simeq \left(\frac{1}{\Omega_b} - \frac{1}{\Omega_b + 4(1+z_{\rm dec}) \Omega_\gamma/3}\right) \delta \Omega_b.
\end{equation}
When using the lensing-potential power spectrum estimates to constrain the CIP amplitude we find that $\Omega_b = 0.0486 \pm0.0014$ at 95\% CL uncertainty and $\Omega_{\gamma} \simeq 10^{-4}$ and $z_{\rm dec} \simeq 1100$. With this we find that the variation in $Q^{TT}(L)$ due to the allowed range of $\Omega_b$ is about 2\%; as a result we evaluate $Q^{TT}(L)$ on the fiducial cosmology stated in the Introduction. Also note that the CIP contribution is nearly independent of the noise properties of the experiment. 
 
\begin{figure}[!ht]
\resizebox{!}{7cm}{\includegraphics{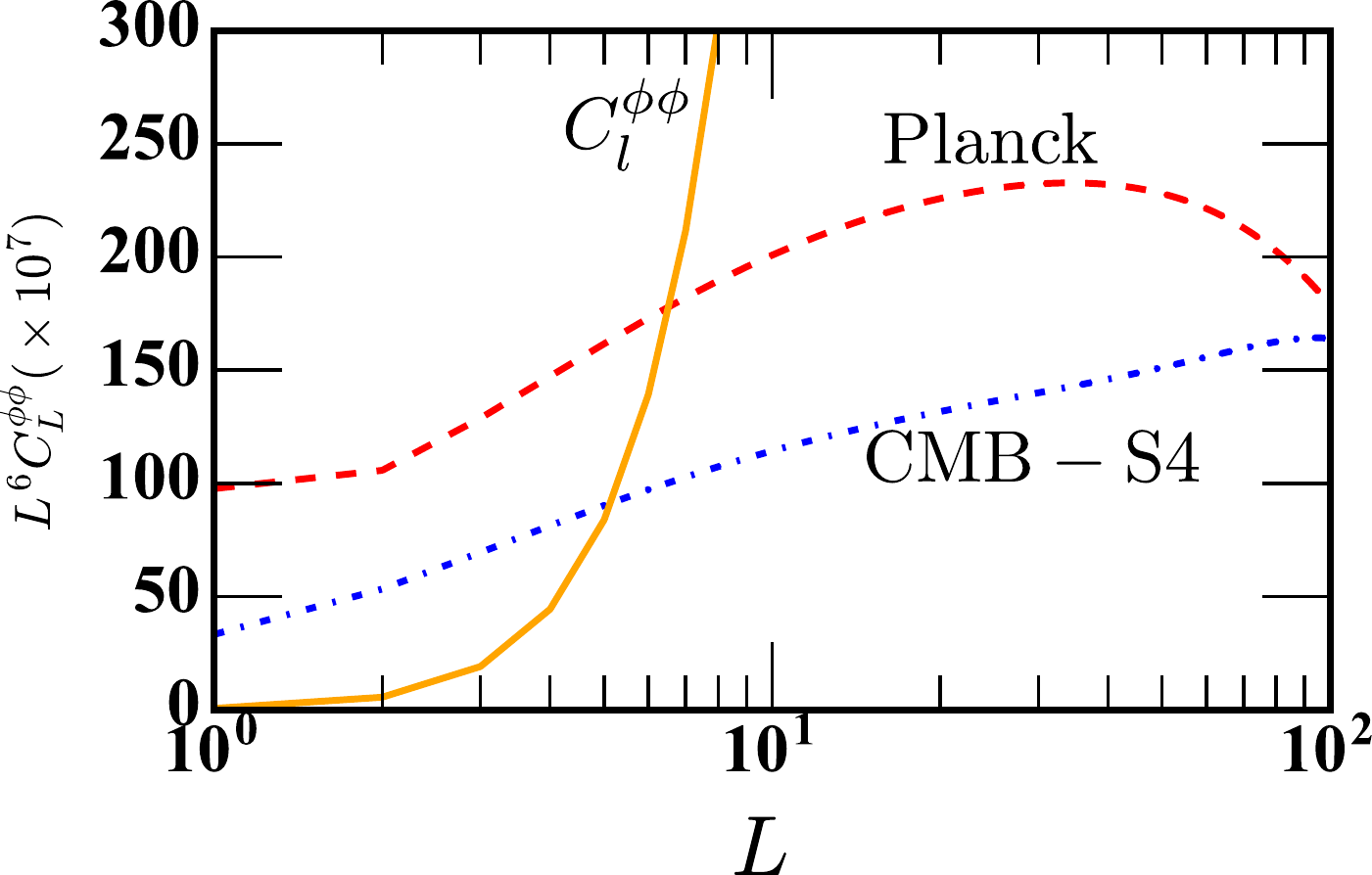}}
\caption{The CIP contribution to the estimator for the lensing potential power spectrum for both Planck (dashed blue) and CMB-S4 (dot-dashed blue) for $\Delta_{\rm rms}^2(R_{\rm CMB}) = 4.34 \times 10^{-3}$. For comparison the lensing potential power spectrum is shown in the solid orange curve.}
\label{fig:CIP_planck_and_S4}
\end{figure}

As noted previously, the lensing potential power spectrum can be estimated from correlations between both temperature and polarization maps of the CMB. The full estimator is an optimally weighted sum of each correlated map, as described in more detail in Appendix \ref{sec:CIPeffects} and \ref{append:fullsky}. The weights are roughly given by the inverse of the Gaussian noise intrinsic to each map.  In particular, Fig.~\ref{fig:clpp_contrib_Planck} shows that for the Planck satellite the $TT$ and $TE$ cross-correlated CMB maps dominate the optimal estimator. 
For a futuristic experiment we consider the CMB Stage-4 (CMB-S4) experiment, which has the noise properties described in Table~\ref{table:plancksens}.
As shown in Fig.~\ref{fig:clpp_contrib_CMBS4}, in this case the $EB$ cross-correlated CMB map dominates the optimal \textit{lensing} estimator.  A similar calculation to that yielding Eq.~(\ref{eq:QTTapprox}) shows that $Q^{EB}(L)$ is suppressed relative to $Q^{TT}(L)$ due to the fact that the numerator of $Q^{EB}(L)$ vanishes at leading order in $L/l_{\rm min}$ (see Appendix  \ref{sec:CIPeffects}):
\begin{eqnarray}
\int\frac{d^2 l_1}{(2\pi)^2}h_{EB}(\vec l_1, \vec l_2)F_{EB}(\vec l_1,\vec l_2) \simeq \int\frac{d^2 l_1}{(2\pi)^2} \cos \varphi_{\vec l_1}\frac{C^{E,dE}_{l_1}}{C_{l_1}^{EE}}L l_1\sin^2 2 \varphi_{\vec l_1 \vec l_2} = 0.
\end{eqnarray}
Since for this idealized experiment the the $EB$ cross-correlated maps dominate the optimal lensing estimator, we expect the overall CIP contribution to future estimates of the lensing power-spectrum to be \textit{less} than that for Planck. As shown in Fig.~\ref{fig:CIP_planck_and_S4}, when this calculation is done on the full sky we find that this is indeed the case with the CIP contribution to Planck about twice as large as it is for CMB-S4. 
Of course, CMB-S4 is significantly more sensitive to CIPs than \textit{Planck} when using the optimal CIP estimator, will as we explain in Sec.~\ref{sec:forecasts}.

We show the first ten values of the CIP contribution to both Planck and CMB-S4 in Table \ref{tab:CIP}. 
	\begin{table*}[hbtp!]
		\begin{tabular}{ l  c  c  }
			\hline
			\hline
			$L$ &  Planck & CMB-S4   \\             
			\hline
			1 & 3.54$\times 10^{-3}$ & 1.20$\times 10^{-3}$   \\
			2 & 5.39$\times 10^{-4}$ & 2.72$\times 10^{-4}$  \\
			3 & 2.30$\times 10^{-4}$ & 1.24$\times 10^{-4}$  \\
			4 & 1.30$\times 10^{-4}$ & 7.20$\times 10^{-5}$  \\
			5 & 8.45$\times 10^{-5}$ & 4.71$\times 10^{-5}$  \\
			6 & 5.93$\times 10^{-5}$ & 3.33$\times 10^{-5}$  \\
			7 & 4.40$\times 10^{-5}$ & 2.48$\times 10^{-5}$  \\
			8 & 3.40$\times 10^{-5}$ & 1.92$\times 10^{-5}$  \\
			9 & 2.70$\times 10^{-5}$ & 1.54$\times 10^{-5}$  \\
			10 & 2.20$\times 10^{-6}$ & 1.25$\times 10^{-6}$  \\
			$>$10 & $(L/0.047)^{-2}$ & $(L/0.029)^{-1.9}$  \\
			\hline
		\end{tabular}
		\caption{CIP contribution to the lensing-potential estimator, $[L(L+1)]^2\hat{C}_L^{\phi \phi}/(2\pi)$, for Planck and CMB-S4.}
		\label{tab:CIP}
	\end{table*}
For $11 \leqslant L \leqslant 40$ both the Planck and CMB-S4 CIP contributions to $[L(L+1)]^2\hat{C}_L^{\phi \phi}/(2\pi)$ are well-fit (to within a few percent) by a power law of the form $(L/L_0)^{-\alpha}$ with $(L_0,\alpha) =( 0.047,2.0)$ for Planck and $(L_0,\alpha) =( 0.029,1.9)$ for CMB-S4. 

\begin{figure}[!ht]
\begin{center}
\resizebox{!}{7cm}{\includegraphics{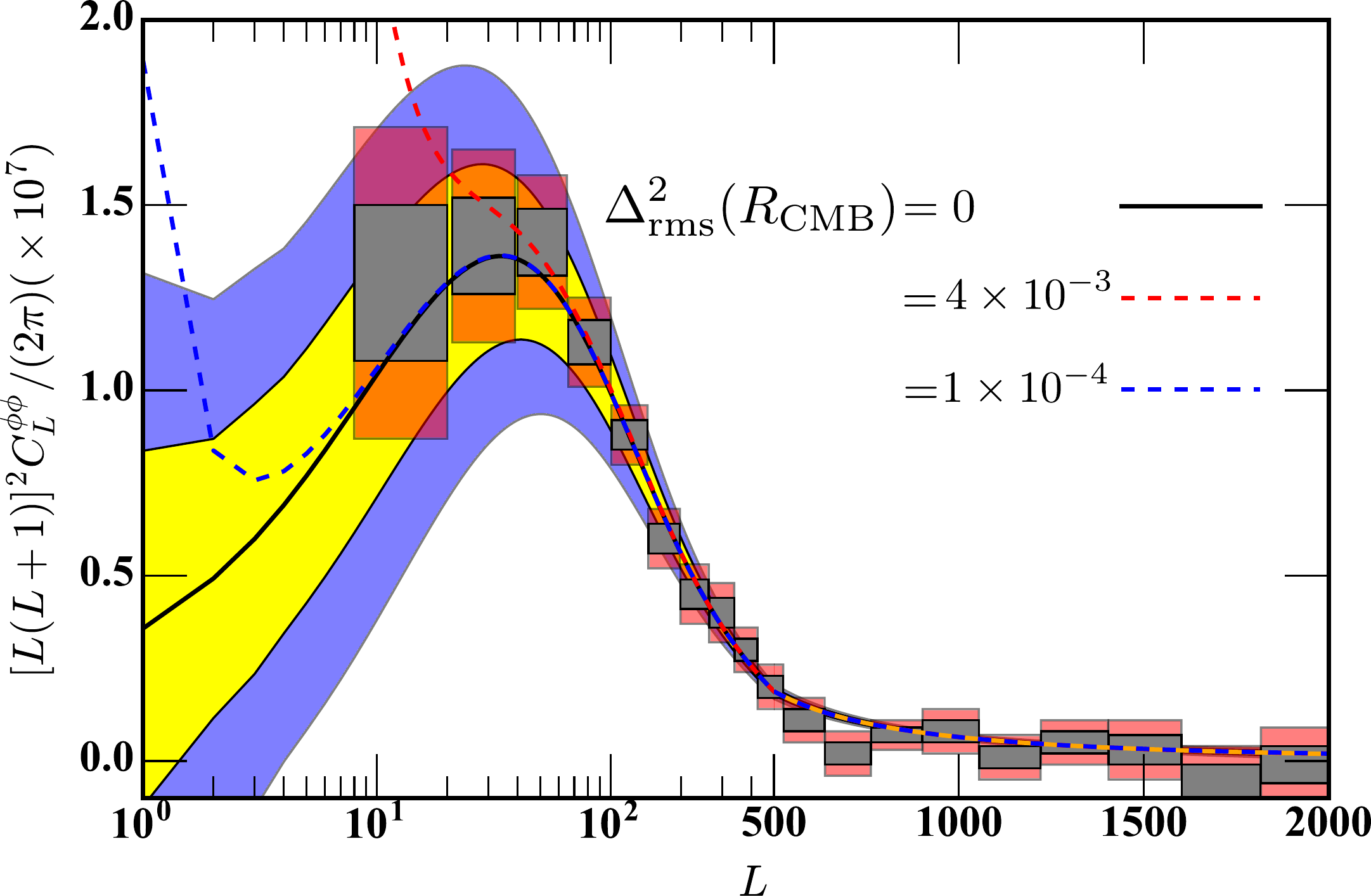}}
\caption{Planck estimates of the lensing potential power spectrum from Ref.~\cite{Ade:2015zua} along with the expectation value of the standard lensing potential power spectrum for three values of $\Delta_{\rm rms}^2(R_{\rm CMB})$. It is clear that the lensing potential estimates should be sensitive to $\Delta_{\rm rms}^2(R_{\rm CMB}) \simeq 0.001$.  We also show the 68\% (yellow region) and 95\% (blue region) CL for a cosmic variance limited estimate of $C_{L}^{\phi \phi}$.}
\label{fig:clpp_wdata}
\end{center}
\end{figure}
Fig.~\ref{fig:clpp_wdata} gives a sense of the sensitivity of the Planck lensing measurements to CIPs. We can see that it is the lowest-$L$ data that provides the constraint on $\Delta_{\rm rms}^2(R_{\rm CMB})$.  Although, in principle the Planck measurements can be extended to even lower $L$-values, in practice estimates of the lensing potential power spectrum are not robust for $L<8$ \cite{Ade:2015zua}.  

\section{Previous constraints to compensated isocurvature perturbations}
\label{sec:oldconstraints}

On scales probed by current cosmological observations (i.e., $k\lesssim 10\ {\rm Mpc}^{-1}$) CIPs lead to a spatial fluctuation of the baryon density beyond the adiabatic prediction. These fluctuations can affect several different observables, such as the primordial light-element abundances and the baryon fraction in galaxies and galaxy clusters \cite{Holder:2009gd}.  It can also lead to anomalous angular correlations in observations of the baryon acoustic oscillations (BAOs). 

Any probe of the spatial dependence of the cosmological baryon fraction has the potential to be sensitive to CIPs.  Past work has concentrated on placing constraints on the CIP amplitude through off-diagonal contributions to the CMB four-point correlation function \cite{Grin:2013uya}, smoothing of the small-scale CMB power spectra \cite{Munoz:2016,Valiviita:2017fbx}, fluctuations in the baryon fraction of galaxy clusters \cite{Holder:2009gd}, spatial variation in the primordial deuterium to hydrogen (D/H) ratio measured in quasar absorption lines \cite{Holder:2009gd}, and an additional offset between the large-scale distribution of total mass from luminous (i.e. baryonic) matter \cite{Soumagnac:2016bjk}.  These constraints probe the CIP modulation on a variety of length scales.  
We will use Eq.~(\ref{eq:DeltaA}) to convert a constraint on the root-mean-square CIP amplitude $\Delta^2_{\rm rms}(R)$ to a constraint on the scale-invariant CIP amplitude, $A_{\rm CIP}$. Since the integral over $k$ formally diverges at small scales we take $k_{\rm min}$ to be of order the current horizon, $k_{\rm min} \simeq (10\ {\rm Gpc})^{-1}$

\subsection{Nucleosynthesis bounds}

The primordial abundance of the light elements are, in part, determined by the local baryon-to-photon ratio, $\eta \equiv n_b/n_\gamma$.  Under the assumption of adiabatic initial conditions, $\eta$ is spatially uniform; in the presence of a CIP it will be modulated by the CIP field, $\Delta$.  Measurements of the primordial helium abundance ($Y_p$) and the D/H ratio in several galaxies allow us to place constraints on any intrinsic scatter in their values.  The measurements of $Y_p$ and D/H place upper limits on $\Delta_{\rm rms}(R_{\rm gal})$, the rms variation of $\Delta$ on galactic scales. 
In the presence of a CIP the Helium abundance, $Y_p$, with vary as $\Delta Y_p \simeq 0.0087 \Delta_{\rm rms}(R_{\rm gal})$ and D/H as $\Delta \log[{\rm D/H}] \simeq 0.69 \Delta_{\rm rms}(R_{\rm gal})$ \cite{Holder:2009gd}.
We will take $R_{\rm gal} \sim 1$ Mpc to be the typical size of the region that collapses to form the galaxies in which $Y_p$ and D/H are measured and find that $\Delta^2_{\rm rms}(R_{\rm gal})  =3.57 A_{\rm CIP}$.  Upper limits to the variation in $Y_p$ and D/H give (at 95\% CL) $\Delta_{\rm rms}(R_{\rm gal})<0.25$ leading to $A_{\rm CIP} < 0.13$. 

\subsection{Baryon-gas fraction bounds}

In the presence of a CIP the ratio $\delta_b/\delta_c$ becomes scale-dependent.  The gas fraction of a galaxy cluster directly probes any fluctuation between baryons and dark matter. Ref.~\cite{Holder:2009gd} found that $\Delta_{\rm rms}(R_{\rm cl}) < 0.08$ at 95\% CL.  With $R_{\rm cl} \simeq 10$ Mpc we have $\Delta^2_{\rm rms}(R_{\rm cl})  =0.49 A_{\rm CIP}$ and $A_{\rm CIP} < 0.017$ at 95\% CL. It is also possible to look for statistical fluctuations in the baryon and dark matter fluctuations. Using tracers of the total matter and luminous matter on large scales ($R \sim 100$ Mpc) Ref.~\cite{Soumagnac:2016bjk} found an upper limit of $A_{\rm CIP} < 0.064$ which translates into $\Delta_{\rm rms}^2(R=100\ {\rm Mpc}) < 0.0165$.

\subsection{Past CMB bounds}

As we saw in Sec.~\ref{sec:cip_cmb}, the CIP modulation smooths the CMB power spectra and contributes to the CMB trispectrum. Constraints from the first effect (under a Gaussian-likelihood approximation \cite{Munoz:2016}) yield the constraint $A_{\rm CIP}<0.04$ at 95\% CL when using both temperature and polarization.  In Ref.~\cite{Grin:2013uya} the CIP contribution to the full trispectrum was constrained using WMAP data, yielding $A_{\rm CIP} < 0.044$ at 95\% CL.  Recently, using temperature and polarization power spectra results from \textit{Planck}, 
the constraint $\Delta_{\rm rms}(R_{\rm CMB})^2 < 0.012$ ($A_{\rm CIP} < 0.050$) at 95\% CL ~\cite{Valiviita:2017fbx}. As discussed in Sec.~\ref{sec:constraints} our power-spectrum only analysis reproduces this result. When using the \textit{Planck} estimates of the lensing power spectrum, Ref.~\cite{Valiviita:2017fbx} neglects to include the CIP contribution to the estimator of the lensing-potential power spectrum. Our full analysis thus improves upon the constraint in Ref.~\cite{Valiviita:2017fbx} by a factor of $\sim 3$.

\section{Constraints to CIPs from \textit{Planck} measurements of the CMB}
\label{sec:constraints}

To explore \textit{Planck}'s sensitivity to CIPs we modified the publicly available Boltzmann solver \texttt{camb}\footnote{\texttt{http://camb.info}} to compute the CIP-modulated CMB power spectra (as described in Appendix \ref{sec:efficient}) and CIP modified expectation value of the standard lensing potential estimator.  We compared these theoretical predictions to the Planck data using the publicly available Planck likelihood code \cite{Aghanim:2015xee} and the Markov Chain Monte Carlo (MCMC) code \texttt{cosmomc}\footnote{\texttt{http://cosmologist.info/cosmomc/}} \cite{Lewis:2002ah}.  
\begin{figure}[!ht]
\begin{center}
\resizebox{!}{8cm}{\includegraphics{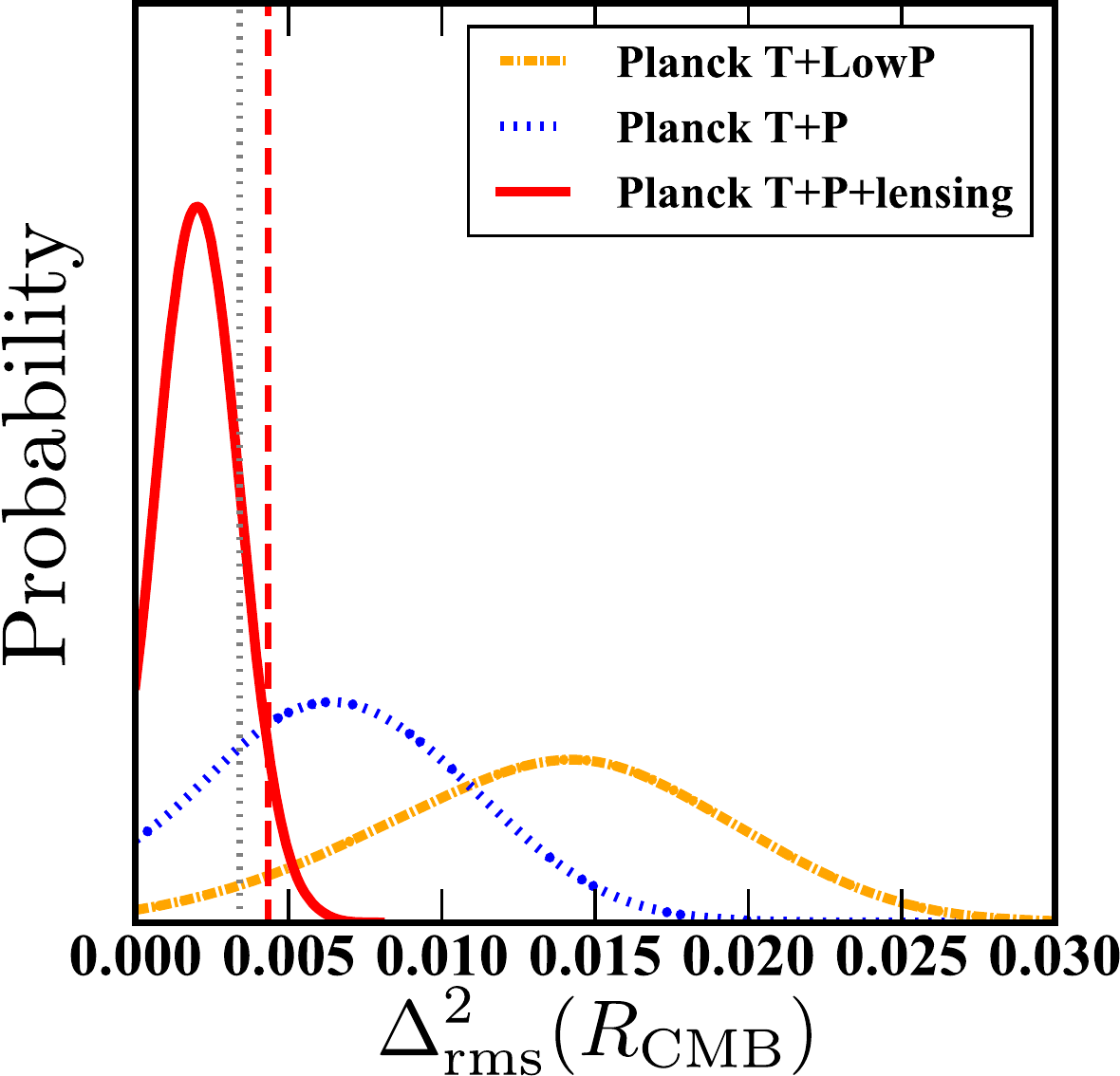}}
\caption{The 1D marginalized posterior for $\Delta^2_{\rm rms}$ using the three combinations of data sets discussed in the text. The vertical dashed  line indicates the 95\% CL upper limit using the T+P+lensing data sets. The vertical dotted line shows the expected 95\% CL upper limit if we were to apply the optimal CIP estimator to the \textit{Planck} measurements of the CMB. The proximity of these two limits shows that, when considering a constraint to the overall amplitude of the CIP modulation, the analysis presented here is nearly optimal. }
\label{fig:CIP1D}
\end{center}
\end{figure}

The \textit{Planck} data has been divided up into a large angular-scale data set (low multipole number) and a small angular-scale data set (high multipole number) \cite{Aghanim:2015xee}.  For all constraints we use the entire range of measurements for the $TT$ power spectrum as well as the low multipole polarization ($TE$ and $EE$) data, which we denote as `T+LowP'.  We also compute constraints using the entire multipole range of polarization measurements, denoted by `T+P'. The division between these two data sets is the multipole number $l =29$ which approximately corresponds to an angular scale of $\simeq 5^\circ$. In addition to the temperature and polarization power spectra we use the \textit{Planck} estimate of the lensing potential power spectrum \cite{Ade:2015zua}.  Since the CIP contribution to the standard lensing estimator is largest on the largest angular scales we use the `aggressive' estimate of the lensing potential power spectrum which extends down to $L_{\rm min} = 8$.  
\begin{table*}[hbtp!]
		\begin{tabular}{ l  c  c  c  c  }
			\hline
			\hline
			Parameter &  T+LowP & T+P & T+P+lensing    \\             
			\hline
			$\omega_b \dots\dots\dots$ & $0.02277 \pm 0.00034$ & 0.02245$\pm$0.00020 & 0.02234$\pm$0.0016  \\
			$\omega_c \dots\dots\dots$ & $0.1166 \pm 0.0025$ & 0.1189$\pm$0.0016 & 0.1186$\pm$0.0014  \\
			$n_s \dots\dots\dots$ & $0.979 \pm 0.0085$ & 0.969$\pm$0.0056 & 0.967$\pm$0.048  \\
			$\log\left( 10^{10}\,A_s \right)$ & $3.061\pm 0.040$& 3.069$\pm$0.036 & 3.04$\pm$0.025  \\
			$\tau \dots\dots\dots..$ & $0.067\pm0.02   $ & 0.068$\pm$0.018 & 0.054$\pm$0.014  \\
			$H_0 \dots\dots\dots$ & $69.2\pm1.3$ & 67.8$\pm$0.75 & 67.83$\pm$0.65  \\
			$\Delta^2_{\rm rms} \dots\dots.$ & $0.014 \pm 0.011   $ & $<$0.0139&$<$0.00434  \\
			\hline
		\end{tabular}
		\caption{Best-fit values and standard deviations for cosmological parameters with the three different Planck data sets as described in the text. 
			All upper limits to $\Delta^2_{\rm rms}$ show 95\% CL. The 68\% CL uncertainty in $\Delta^2_{\rm rms}$ is $\pm 0.0055$ when using T+LowP.}
		\label{tab:constraints}
	\end{table*}

As demonstrated in Fig.~\ref{fig:Planck_diff} polarization data can break degeneracies present in a temperature-only analysis.  The top panel in Fig.~\ref{fig:Planck_diff} shows that the residual $TT$ power spectrum has an oscillating structure around $l \simeq 1000$.  When comparing the CIP modulation of the $TT$ power spectrum we can see by eye that a non-zero CIP amplitude can fit these residuals.  As shown in the first column of Table \ref{tab:constraints} this is reflected by the fact that the the T + LowP data sets prefer a non-zero CIP modulation with $\Delta_{\rm rms}^2(R_{\rm CMB}) = 0.014 \pm 0.011$ at 95\% CL ($A_{\rm CIP}=0.056 \pm 0.044$). When we include the full polarization measurements from \textit{Planck} (i.e., `T+P') the evidence for this CIP mode decreases to $< 0.0139$ at 95\% CL ($A_{\rm CIP}<0.056$). Fig.~\ref{fig:Planck_diff} indicates this is mainly due to precise measurements of $C_{l}^{EE}$ at multipoles between $30 \lesssim l \lesssim 400$. Finally, when including estimates of the lensing potential power spectrum (i.e., `T+P+lensing') the evidence for the CIP modulation decreases by a factor of $\sim$3.2.

The 1D marginalized posterior on $\Delta_{\rm rms}^2(R_{\rm CMB})$ using the three combination of data sets is shown in Fig.~\ref{fig:CIP1D}. We can see how the CIP contributions to the lensing potential power spectrum significantly increases the sensitivity to $\Delta_{\rm rms}^2(R_{\rm CMB})$. This figure also makes clear how the maximum of the 1D posterior decreases as we add in additional data. 

\begin{figure}[!ht]
\begin{center}
\resizebox{!}{17cm}{\includegraphics{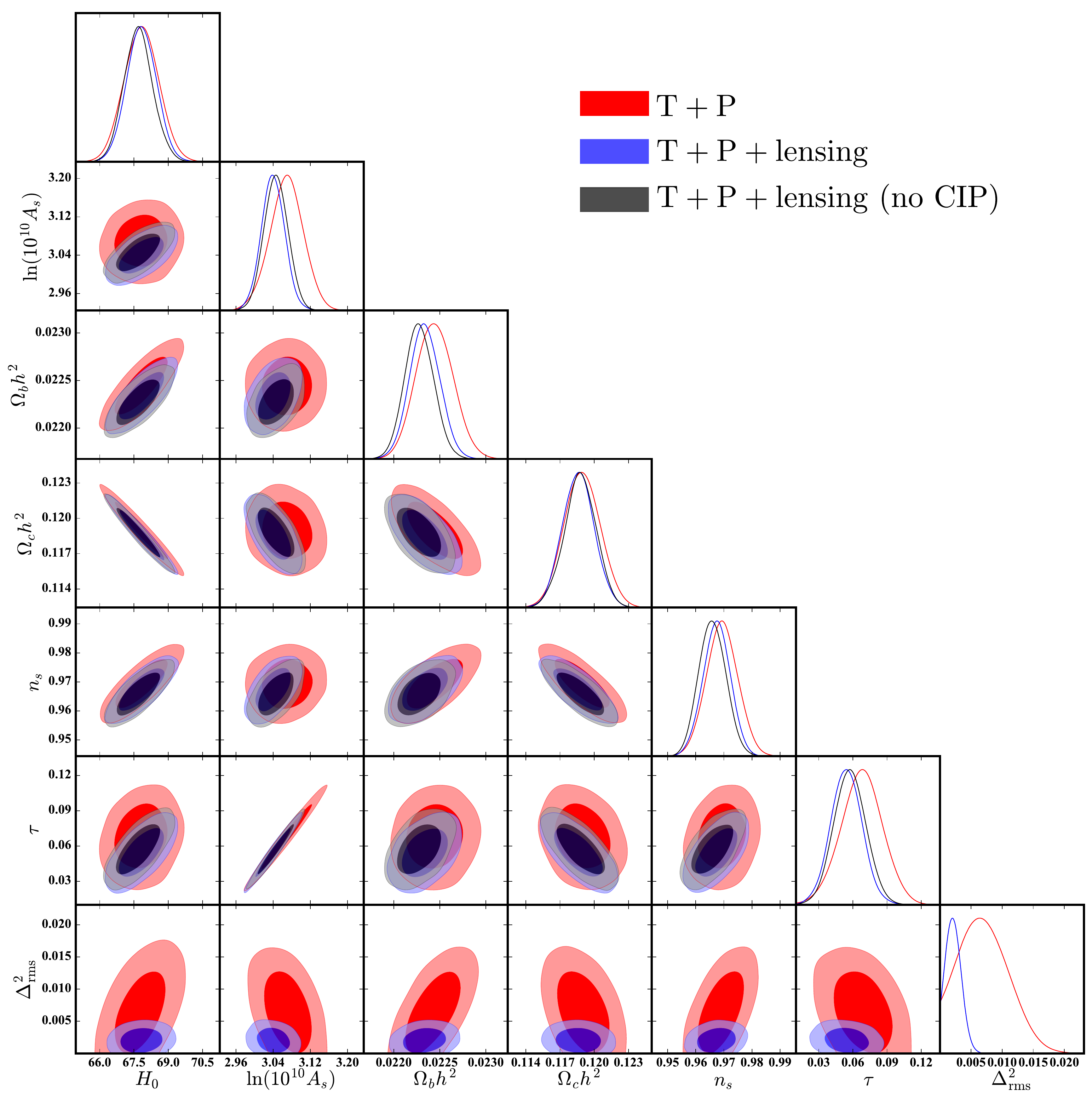}}
\caption{A `triangle-plot' showing the 2D marginalized posteriors for the standard six cosmological parameters and the CIP rms variation on cosmological scales, $\Delta_{\rm rms}^2(R_{\rm CMB})$. }
\label{fig:TriCla}
\end{center}
\end{figure}

In order to explore any degeneracies in these parameters we show a `triangle' plot for the standard six cosmological parameters and $\Delta_{\rm rms}^2(R_{\rm CMB})$ in Fig.~\ref{fig:TriCla}. As shown by a comparison between the black and blue contours, the constraints on the standard six parameters remain fairly unchanged when we additionally constrain $\Delta_{\rm rms}^2(R_{\rm CMB})$. 
Moreover, going from T+P to T+P+lensing clearly reduces the overall uncertainties of $\Delta_{\rm rms}^2(R_{\rm CMB})$, as well as its correlations with the rest of $\Lambda$CDM parameters, since the information on the CIPs comes from the 
large-scale lensing estimator.

\section{Future constraints}
\label{sec:forecasts}

We now assess the sensitivity of future CMB experiments to CIPs. As for lensing, the improvements will come primarily come from small-scales (in particular, polarization), and so we focus on the proposed CMB-S4 experiment, as described in Ref.~\cite{1610.02743}; CMB-S4 will be a nearly cosmic-variance-limited (CVL) experiment in both temperature/polarization. Our analysis can be easily implemented using noise specifications for other future experiments.

We model CMB-S4 as a single channel experiment, as described in Table~\ref{table:plancksens}, observing in the range $l=30\to 3000$ for temperature and $l=30 \to 5000$ for polarization (due to the smaller relative amplitude of small-scale polarized foregrounds). Given these characteristics the lensing noise can be computed as in Refs.~\cite{Okamoto:2003zw,1010.0048} (we show the CVL case in Fig.~\ref{fig:clpp_wdata}; the CMB-S4 estimate still has reconstruction noise due to sample variance).

\subsection{Lensing potential bias and power-spectrum smoothing}

\begin{figure}[!ht]
\begin{center}
\resizebox{!}{7cm}{\includegraphics{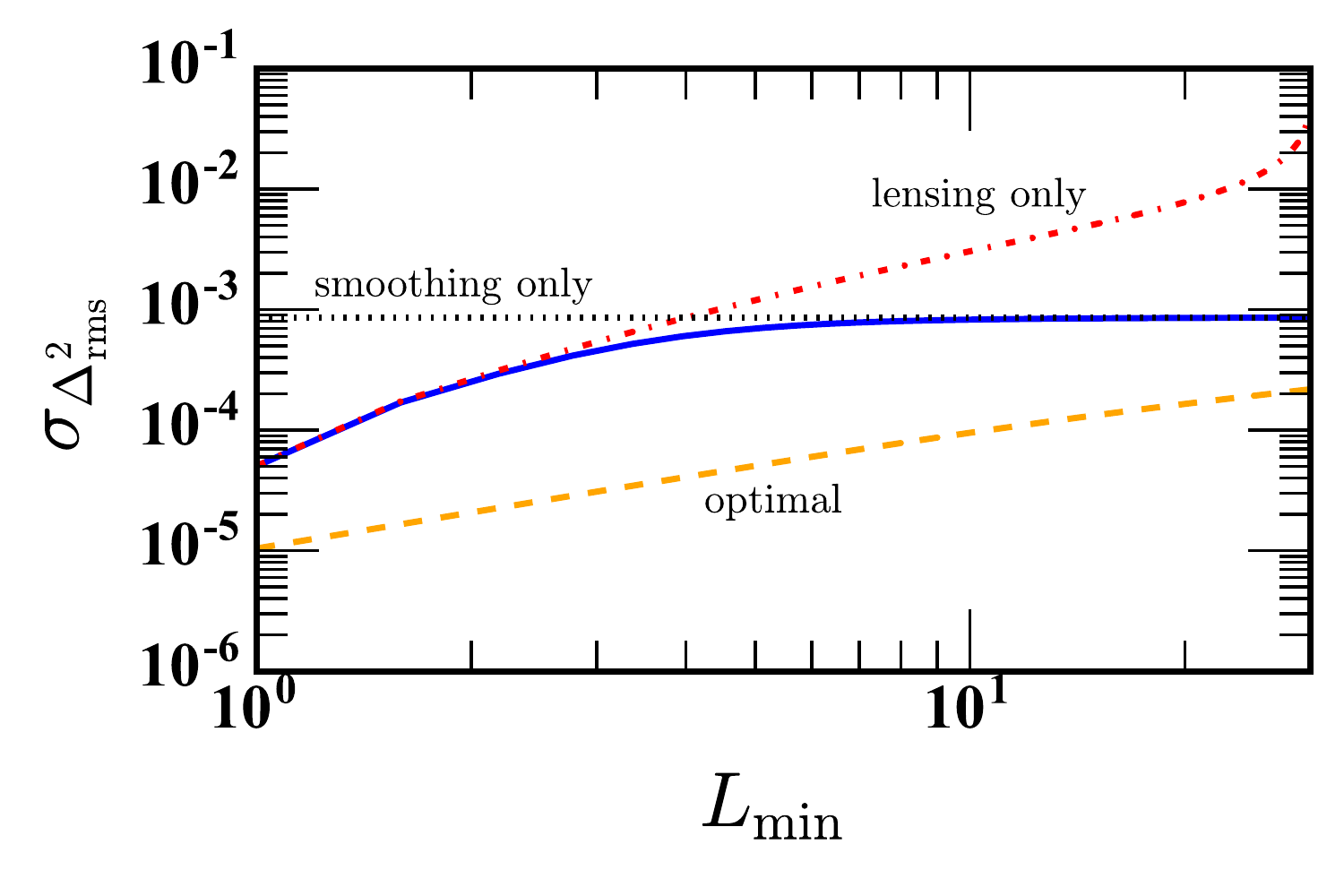}}
\caption{Projected sensitivity to $\sigma_{\rm rms}^2$ for CMB S4. The overall sensitivity is shown in the solid blue curve. The sensitivity can be divided into a contribution from the smoothing of the CMB multipoles (dotted black) and from estimates of the lensing potential power spectrum (dot-dashed red).  The lensing potential contribution is only important if future experiments can probe $L_{\rm min} \lesssim 5$.}
\label{fig:lvstri}
\end{center}
\end{figure}

The sensitivity of lensing potential bias and power-spectrum smoothing to CIPs is estimated using the Fisher matrix \cite{astro-ph/9512139,astro-ph/9611125,astro-ph/9609169,astro-ph/9609170,1402.4108}
\begin{equation}
F_{ij} = \sum_l \dfrac{2l+1}{2} f_{\rm sky} {\rm Tr} \left [ \mathbf C_l^{-1} \dfrac{\partial\mathbf C_l}{\partial p_i}\mathbf C_l^{-1} \dfrac{\partial\mathbf C_l}{\partial p_j} \right],
\end{equation}
where $p_i$ contains the six $\Lambda$CDM parameters plus the CIP variance $\Delta^2_{\rm rms}$, and the CMB covariance matrix is given by
\begin{align}
    \mathbf C_l = \begin{pmatrix} C_l^{TT, {\rm t}} &  C_l^{TE, {\rm t}} &  C_l^{Td, {\rm t}}\\  C_l^{TE, {\rm t}} & C_l^{EE, {\rm t}} & 0
    \\  C_l^{Td, {\rm t}} & 0 & C_l^{dd, {\rm t}}\end{pmatrix},
\end{align}
where all of the power spectra are computed from the observed maps as in Eqs.~(\ref{eq:obsmaps}, \ref{eq:knoxknoise}), following the noise properties from Table~\ref{table:plancksens} with zero noise for any cross-correlated (i.e., $TE$) maps.
Since the CIP contribution to the lensing-potential power-spectrum estimator roughly decreases as $L^{-2}$, the sensitivity of future estimates to the scale-invariant CIP amplitude, $A_{\rm CIP}$, is highly dependent on the minimum observable $L$-value, $L_{\rm min}$, and therefore highly dependent on the sky coverage.  Assuming that the non-lensing biases contributing to the lensing potential power spectrum estimator can be robustly subtracted on large angular scales, the minimum multipole which can be estimated is approximately given by $L_{\rm min} \sim f_{\rm sky}^{-1/2}$. Unfortunately, galactic foregrounds \cite{1405.7351} and temperature/polarization leakage \cite{1610.02743} could degrade the largest-scale measurements ($L<30$). 

We defer analysis of these complications to future work and obtain forecasts as a function of the minimum $L_{\rm min}$ detectable by the CMB S4. Using information from the lensing estimator and power-spectrum smoothing, we obtain the sensitivity to $\Delta_{\rm rms}^2(R_{\rm CMB})$ for CMB-S4 experiment as a function of $L_{\rm min}$ (shown in Fig.~\ref{fig:lvstri}). For $L_{\rm min}\geq 10$ the majority of the constraint comes from the CIP modulation of the CMB power spectrum, whereas for $L_{\rm min}<10$ the CIP contribution to the lensing potential estimator dominates. We note that the CMB-S4 lensing noise is very close to the cosmic-variance limit, and therefore the results should be the same for any other nearly CVL experiment. In particular, a full-sky CVL measurement of the lensing potential power spectrum has the potential to constrain $\Delta_{\rm rms}^2(R_{\rm CMB}) \lesssim 10^{-5}$, and therefore $A_{\rm CIP}\lesssim 4\times 10^{-5}$. We now turn our attention to additional improvements possible with an optimal estimator of CIPs.

\subsection{Optimal estimator}

The constraints obtained in this work from the observed CMB trispectrum rely on the contribution of CIPs to the standard lensing potential estimator. There is, however, an optimal CIP estimator which relies on the distinct (from lensing) off-diagonal CMB multipole correlations induced by CIPs, as shown in Refs. \cite{Grin:2011tf,Grin:2011nk,He:2015msa} and summarized in Appendices \ref{append:fullsky}-\ref{append:optimal}. This estimator was used to obtain the WMAP constraints to CIPs in Ref.~\cite{Grin:2013uya}. The Fisher information $F$, which yields the minimum uncertainty $\sigma_{\Delta_{\rm rms}^{2}}=\sqrt{1/F}$ is given by
\begin{eqnarray}
F=\sum_{L}\frac{\left(2L+1\right)}{2}f_{\rm sky}\left(\frac{\partial C_{L}^{\Delta \Delta}}{\partial \Delta_{\rm rms}^{2}}   \right)^{2}\left(N_{L}^{\Delta \Delta}\right)^{-2}, \label{eq:fisher_optimal}
\end{eqnarray} 
where $N_{L}^{XX'}$ are defined in Ref.~\cite{Grin:2013uya} and computed under the null hypothesis. We use Eq.~(\ref{eq:fisher_optimal}) to forecast the sensitivity of \textit{Planck} and CMB-S4  (although \textit{Planck} data are public, we still have to `forecast' its sensitivity given that a full analysis of the \textit{Planck} CMB trispectrum for CIPs does not yet exist) to a scale-invariant angular power spectrum of CIPs.

For \textit{Planck} noise parameters, we find that the optimum estimator has a $2\sigma$ sensitivity of $\Delta_{\rm rms}^{2}(R_{\rm CMB})\simeq 4 \times 10^{-3}$, offering no significant improvement over the constraint from the CIP contribution to the lensing potential estimator (see Fig.~\ref{fig:CIP1D} for an illustration). In other words, the constraints in Sec.~\ref{sec:constraints} are nearly optimal using \textit{Planck} data.

On the other hand, for a nearly CVL experiment like CMB-S4, the optimal estimator improves on the constraint from the CIP contribution to the lensing estimator by a factor of $\sim 4.0$ for $L_{\rm min}=1$ and a factor of $\sim 4.5$ for $L_{\rm min}=30$. This dramatic difference, illustrated in Fig.~\ref{fig:lvstri}, is driven by the constraining power of a nearly CVL polarization experiment.

Given the fact that the trispectrum is so much more constraining than the CIP-induced smoothing of the CMB power spectrum, we neglect primary power-spectrum constraints in this Fisher analysis. For futuristic experiments (like CMB-S4), the reconstruction noise for both lensing and CIPs may be low enough that lensing could introduce a significant bias \cite{Heinrich:2016gqe} to the estimators described in Appendix \ref{append:optimal}, requiring either a debiased minimum-variance estimator (as discussed in Ref.~\cite{Su:2011ff}) or a `delensed' CMB map (as discussed in Refs.~\cite{Smith:2008an,Smith:2010gu,Larsen:2016wpa}), in which lensing-induced correlations have been filtered out. We defer an analysis that includes these complications to future work, and simply note that Eq.~(\ref{eq:fisher_optimal}) quantifies the best CIP reconstruction we could ever achieve using the CMB.

\section{Conclusions}
\label{sec:conclusions}

We have shown how the presence of a CIP modulation leads to additional contributions to the standard lensing potential power spectrum estimator applied to CMB maps.  In particular we have used the \textit{Planck} data to place the most stringent constraints on the amplitude of a scale-invariant CIP modulation, $\Delta_{\rm rms}^2(R_{\rm CMB})<4.3\times 10^{-3}$ at 95\% CL, on cosmological scales. The method discussed here provides a nearly optimal upper limit when using \textit{Planck} data. We note that this statement only applies to the overall amplitude of a scale-invariant CIP power spectrum. The optimal analysis would additionally provide constraints to the scale-dependent CIP power spectrum at different multipoles. 
\begin{table}[!ht]
\begin{center}
\noindent\adjustbox{max width=1.2\textwidth}{%
\begin{tabular}{p{60mm}l}
\hline\hline
Method & $A_{\rm CIP}$ \\\noalign{\smallskip}\hline\noalign{\smallskip}
Trispectrum, WMAP  & $< 0.044$\\\noalign{\smallskip}
Baryon Acoustic Oscillations, SDSS-III   & $<0.064$\\\noalign{\smallskip}
Baryon Fraction in Galaxy Clusters   &$<0.017$ \\\noalign{\smallskip}
Dispersion in $^4$He and D/H  &$< 0.13$ \\\noalign{\smallskip}\hline
\noalign{\smallskip}
\emph{This work}: & \\\noalign{\smallskip}
Planck T+LowP  & $0.056 \pm 0.044$ \\\noalign{\smallskip}
Planck T+P   & $<0.056$ \\\noalign{\smallskip}
Planck T+P+lensing   & $<0.017$ 
\\\noalign{\smallskip}
\hline
\end{tabular}}
\caption{Current constraints on $A_{\rm CIP}$, the amplitude of the power spectrum of a scale-invariant CIP modulation field.  All uncertainties and upper limits are at 95\% CL.  \label{table:cips}} 
\end{center}
\end{table}

We show a comparison of our results to previous constraints to the amplitude scale-invariant CIP power spectrum in Table \ref{table:cips} [see Eq.~(\ref{eq:DeltatoACIP}) for a translation between $\Delta_{\rm rms}^2(R_{\rm CMB})$ and $A_{\rm CIP}$].  This table shows that before this work the most sensitive constraint to the CIP amplitude came from estimates of the baryon fraction in galaxy clusters. The results presented here are as sensitive to $A_{\rm CIP}$. Furthermore, given that the analysis of the baryon fraction estimates assumes that the clusters are representative of the baryon density throughout the universe and that they are kinematically relaxed \cite{Holder:2009gd} the robustness of these constraints may be in question \cite{Munoz:2016}.  On the other hand the CMB-related constraints presented here do not suffer from such complexities and are therefore easier to interpret.  

We note that the constraints presented here do not apply to curvaton-generated CIPs, which are correlated with adiabatic fluctuations. In this scenario, the effect of CIPs on CMB observables is enhanced relative to what is considered here. Our limits do, however, give us a conservative upper limit on the sensitivity of our technique to curvaton-generated CIPs.

As discussed in Sec.~\ref{sec:constraints}, the upper limit to $A_{\rm CIP}$ found here is nearly optimal for the measurements from \textit{Planck}.  Looking towards future experiments with nearly ideal sensitivity to polarization, we find that the method presented here is about a factor of 4.5 less sensitive than the optimal estimator.  However, we note that the standard analysis of future CMB data will include an estimate of the lensing potential power spectrum from the CMB trispectrum.  Therefore, the results presented here show that, in the presence of a CIP modulation, we may find excess power in the lensing estimator at large angular scales.  This conclusion applies to \emph{any} scale-invariant stochastic field which modulates the CMB.  In particular, processes like patchy reionization \cite{2013PhRvD..87d7303G} or a spatial variation in the fine structure constant \cite{Sigurdson:2003pd,King:2012id}, should contribute to the lensing estimator in a similar way.  We leave such extensions of the results presented here to future work. 

\acknowledgments

TLS acknowledges support from the Provost's office at Swarthmore College. 
JBM is supported by the Simons foundation and NASA ATP Grant No.~NNX15AB18G.
RS acknowledges the support of the Panaphil Foundation though a Velay Scholarship. 
The authors wish to thank Yacine Ali-Ha\"imoud, Marc Kamionkowski, Michael Kesden, and Marius Millea for 
interesting and enlightening discussions.

\begin{appendix}

\section{CIP modulation of the CMB on the flat sky}
\label{sec:CIPeffects}

In the presence of a CIP field the CMB is modulated in a way that is analogous to the effects of gravitational lensing.  In this appendix we present the flat-sky expressions for the modulated temperature and polarization. Since we are interested in the leading order effects on the trispectrum we only consider terms to linear order in the lensing potential, $\phi$, and CIP modulation field, $\Delta$. 

In the flat-sky approximation we have \cite{Hu:2001kj, Okamoto:2003zw,PhysRevD.67.123507}
\begin{eqnarray}
(Q\pm U)(\hat n) &=& - \int \frac{d^2 l}{(2\pi)^2} \left[ E(\vec l) \pm i B(\vec l)\right] e^{\pm 2 i \varphi_{\vec l}} e^{i \vec l \cdot \hat n},
\end{eqnarray}
where $\varphi_{\vec l}$ is the polar angle of the wave vector $\vec l$.  From this it is straightforward to show that 
\begin{eqnarray}
E(\vec l) &=& - \frac{1}{2} \left( \cos 2 \varphi_{\vec l} \left[Q+U\right](\vec l) + 
\sin 2 \varphi_{\vec l} \left[Q-U\right](\vec l)\right), \\
B(\vec l) &=& \frac{1}{2i} \left( \sin 2 \varphi_{\vec l} \left[Q+U\right](\vec l) - \cos 2 \varphi_{\vec l} 
\left[Q-U\right](\vec l) \right),
\end{eqnarray}
where $\left[Q\pm U\right](\vec l) = \int d^2 n \left[Q\pm U\right](\hat n) e^{- i \vec l \cdot \hat n}$. Expanding $\left[Q\pm U\right](\hat n)$ to linear order in the lensing potential and the CIP modulation field we have 
\begin{equation}
\left[Q\pm U\right](\hat n+\vec \nabla \phi, \Delta) \simeq \left[Q\pm U\right](\hat n) + \nabla_i \phi \nabla^i \left[Q\pm U\right] + \Delta(\hat n) \frac{\partial \left[Q\pm U\right]}{\partial \Delta} \bigg|_{\Delta = 0}.
\end{equation}
From this we can write 
\begin{eqnarray}
\left[Q+ U\right]^{\rm obs}(\vec l) &=& \left[Q+ U\right]^{(0)}(\vec l) + \int \frac{d^2 l_1}{(2\pi)^2}\bigg[ W_E(\vec l, \vec l_1) \cos 2\varphi_{\vec l_1}-W_B(\vec l, \vec l_1)\sin 2 \varphi_{\vec l_1}\bigg],\\
\left[Q-U\right]^{\rm obs}(\vec l) &=& \left[Q- U\right]^{(0)}(\vec l) + \int \frac{d^2 l_1}{(2\pi)^2}\bigg[W_B(\vec l, \vec l_1) \cos 2\varphi_{\vec l_1}\nonumber -W_E(\vec l, \vec l_1)\sin 2 \varphi_{\vec l_1}\bigg],
\end{eqnarray}
where 
\begin{eqnarray}
W_X(\vec l, \vec l_1) &\equiv& X(\vec l_1)W_\phi(\vec l_1, \vec l) + \frac{\partial X(\vec l_1)}{\partial \Delta} \Delta(\vec l-\vec l_1)+ \frac{1}{2}\frac{\partial^2 X}{\partial \Delta^2} \int \frac{d^2 l_1'}{(2\pi)^2} \Delta(\vec l_2') \Delta(\vec l - \vec l_1-\vec l_1'),\\
W_\phi(\vec l, \vec l_1) &\equiv& -\phi(\vec l- \vec l_1) [(\vec l- \vec l_1) \cdot \vec l_1]-\frac{1}{2} \int \frac{d^2 l_1'}{(2\pi)^2} \phi(\vec l_1') \phi(\vec l- \vec l_1 - \vec l_1')\left[\vec l_1' \cdot \vec l_1\right]\left[(\vec l_1'+ \vec l_1 - \vec l)\cdot \vec l_1\right] .
\end{eqnarray}
Writing $X_{\rm obs}(\vec l) = X(\vec l) + \delta X(\vec l)$ we have 
\begin{eqnarray}
\delta T(\vec l) &=& \int \frac{d^2 l_1}{(2\pi)^2} W_T(\vec l, \vec l_1), \\
\delta E(\vec l) &=& \int \frac{d^2 l_1}{(2\pi)^2} \left[W_E(\vec l, \vec l_1) \cos 2 \varphi_{\vec l_1\vec l} - 
W_B(\vec l, \vec l_1) \sin 2 \varphi_{\vec l_1 \vec l}\right] , \\
\delta B(\vec l) &=& \int \frac{d^2 l_1}{(2\pi)^2} \left[W_B(\vec l, \vec l_1) \cos 2 \varphi_{\vec l_1\vec l} + 
W_E(\vec l, \vec l_1) \sin 2 \varphi_{\vec l_1 \vec l}\right], \\
\end{eqnarray}
where $\varphi_{\vec l_1, \vec l} \equiv \varphi_{\vec l_1} - \varphi_{\vec l}$. 
This allows us to write down an estimator for the lensing potential in terms of the various correlations \cite{Hu:2001kj, Okamoto:2003zw,PhysRevD.67.123507}:
\begin{equation}
\hat d_{\alpha}(\vec L) \equiv \frac{i \vec L A_{XX'}(L)}{L^2} \int \frac{d^2 l_1}{(2\pi)^2} \frac{1}{2} \left[X^t(\vec l_1) X^{'t}(\vec l_2) + X^{'t}(\vec l_1) X^{t}(\vec l_2)\right]F_{XX'}(\vec l_1, \vec l_2),
\end{equation}
where $\alpha = XX'$ and 
\begin{eqnarray}
F_{XX'}(\vec l_1, \vec l_2) &\equiv& \frac{C_{l_1}^{X'X',{\rm t}} C_{l_2}^{XX,{\rm t}} f_{XX'}(\vec l_1, \vec l_2) - C_{l_1}^{XX',{\rm t}} C_{l_2}^{XX',{\rm t}} f_{XX'}(\vec l_2,\vec l_1)}{C_{l_1}^{XX,{\rm t}} C_{l_2}^{X'X',{\rm t}} C_{l_1}^{X'X',{\rm t}} C_{l_2}^{XX,{\rm t}} - (C_{l_1}^{XX',{\rm t}} C_{l_2}^{XX',{\rm t}})^2},\\ 
A_{XX'}(L) &\equiv& L^2 \left\{ \int \frac{d^2 l_1}{(2\pi)^2} \frac{1}{2} \left[f_{XX'}(\vec l_1, \vec l_2) + 
f_{XX'}(\vec l_2, \vec l_1)\right] F_{XX'}(\vec l_1,\vec l_2)\right\}^{-1}.
\end{eqnarray}
The (Gaussian) noise in each estimator is given by 
\begin{equation}
N^{(0)}_{\alpha,\beta}(L) = \frac{A_\alpha(L) A_{\beta}(L)}{L^2} \int \frac{d^2 l_1}{(2\pi)^2} F_{\alpha}(\vec l_1, \vec l_2) \left[F_{\beta}(\vec l_1, \vec l_2) C_{l_1}^{XY,{\rm t}} C_{l_2}^{X'Y',{\rm t}}+F_{\beta}(\vec l_2, \vec l_1) C_{l_1}^{XY',{\rm t}} C_{l_2}^{X'Y,{\rm t}}\right],
\end{equation}
with $\beta = (YY')$. 
Finally, we can combine all of these to form a `minimum variance' estimator \cite{Hu:2001kj, Okamoto:2003zw}
\begin{equation}
\hat d_{\rm MV}(\vec L) \equiv \sum_{\alpha} w_\alpha(L) \hat{d}_{\alpha}(\vec L),
\end{equation}
where the weights $w_\alpha$ are chosen to minimize the Gaussian noise, 
\begin{equation} 
w_\alpha(L) \equiv \frac{\sum_\beta ({\bf N}^{-1})_{\alpha \beta}}{\sum_{\beta \gamma} ({\bf N}^{-1})_{\beta\gamma}}.
\end{equation}

\begin{table}[!ht]
\begin{tabular*}{11cm}{@{\extracolsep{\fill}}ccc}
\\\noalign{\smallskip}\hline\noalign{\smallskip}
 $XX'$&$f_{ X X'}(\vec l_1, \vec l_2)$&$h_{ X X'}(\vec l_1, \vec l_2)$\\\noalign{\smallskip}\hline\noalign{\smallskip}
TT &$ C_{l_1}^{TT}(\vec L \cdot \vec l_1) + C_{l_2}^{TT} (\vec L \cdot \vec l_2)$ & $C_{l_1}^{T,dT} + C_{l_2}^{T,dT}$ \\\noalign{\smallskip}
TE&$ C_{l_1}^{TE} \cos 2\varphi_{\vec l_1 \vec l_2} (\vec L \cdot \vec l_1) + C_{l_2}^{TE}(\vec L \cdot \vec l_2)$ & $ C_{l_1}^{T,dE} \cos2 \varphi_{\vec l_1 \vec l_2}  + C_{l_2}^{E,dT}$  \\ \noalign{\smallskip}
TB&$ C_{l_1}^{TE} \sin 2 \varphi_{\vec l_1 \vec l_2} (\vec L \cdot \vec l_1)$ & $ C_{l_1}^{T,dE} \sin 2 \varphi_{\vec l_1 \vec l_2} $  \\ \noalign{\smallskip}
EE&$ \left[C_{l_1}^{EE} (\vec L \cdot \vec l_1) + C_{l_2}^{EE}(\vec L \cdot \vec l_2)\right] \cos 2 \varphi_{\vec l_1 \vec l_2} $ & $ \left[C_{l_1}^{E,dE}  + C_{l_2}^{E,dE}\right] \cos 2 \varphi_{\vec l_1 \vec l_2} $ \\ \noalign{\smallskip}
EB &$\left[C_{l_1}^{EE}(\vec L \cdot \vec l_1) - C_{l_2}^{BB} (\vec L \cdot \vec l_2) \right] \sin2 \varphi_{\vec l_1 \vec l_2}$ & $\left[C_{l_1}^{E,dE} - C_{l_2}^{B,dB}  \right] \sin2 \varphi_{\vec l_1 \vec l_2}$   \\ \noalign{\smallskip}
BB&$ \left[C_{l_1}^{BB}(\vec L \cdot \vec l_1) + C_{l_2}^{BB} (\vec L \cdot \vec l_2)\right] \cos 2 \varphi_{\vec l_1 \vec l_2} $ & $ \left[C_{l_1}^{B,dB} + C_{l_2}^{B,dB} \right] \cos 2 \varphi_{\vec l_1 \vec l_2} $ \\\noalign{\smallskip}\hline\noalign{\smallskip}
\end{tabular*}
\caption{The lensing and CIP trispectrum response functions. } 
\label{tab:lens_resp}
\end{table}

Analogous to the effects of lensing, in the presence of CIPs the bias-subtracted lensing potential power spectrum estimator has an expectation value \cite{Okamoto:2003zw,PhysRevD.67.123507}
\begin{equation}
L^2 \langle \hat{C}^{\phi \phi}_L\rangle = L^2 C_L^{\phi \phi} + L^2 C_L^{\Delta \Delta} \sum_{\alpha,\beta} w_\alpha(L)w_\beta(L)Q_{\alpha}(L)Q_{\beta}(L),
\end{equation}
where
\begin{equation}
Q_\alpha(L) \equiv \frac{\int\frac{d^2 l_1}{(2\pi)^2}[h_{XX'}(\vec l_1, \vec l_2)+h_{X'X}(\vec l_1, \vec l_2)]F_{XX'}(\vec l_1,\vec l_2) }{\int \frac{d^2 l_1}{(2\pi)^2} [f_{XX'}(\vec l_1, \vec l_2)+f_{X'X}(\vec l_1, \vec l_2)] F_{XX'}(\vec l_1,\vec l_2)}.
\label{eq:genex}
\end{equation}
Note that we have left off the non-gaussian and residual CIP gaussian bias from this expression. As shown in Sec.~\ref{sec:cip_cmb}, given the current upper limits to $A_{\rm CIP}$ these terms make a negligible contribution to the expectation value of the temperature-only estimator.  In additional to this, Ref.~\cite{PhysRevD.67.123507} demonstrates that the contribution to these biases from correlations other than $TT$ are of about the same order of magnitude and do not change significantly with an improvement in the instrumental noise.  

\section{Full sky expressions}
\label{append:fullsky}

The Planck CMB lensing analysis in Ref.~\cite{Ade:2015zua} used the full-sky estimators derived in Ref.~\cite{Okamoto:2003zw} to compute the lensing potential power spectrum.  In this Appendix we compute the dominant CIP contribution to that estimator. 

Consider an ensemble of CMB fields modulated by both a fixed deflection field $\phi$ and a fixed CIP field $\Delta$.  These second order effects produce an off-diagonal covariance \cite{Okamoto:2003zw,Grin:2011nk,Grin:2011tf} 
\be
\VEV{X_{l m} X'_{l'm'}}\big|_{\rm lens,CIP} = C_l^{X X'} \delta_{l l'} \delta_{m m'} (-1)^m + 
\sum_{LM} (-1)^M \wigner l m {l'} {m'} L {-M} \left[\phi_{LM} f^{XX'}_{l Ll'}+ \Delta_{LM} h^{XX'}_{l L l'}\right],
\label{eq:2pt}
\ee
where $\phi_{LM}$ and $\Delta_{LM}$ are the multipoles of the lensing potential and CIP field respectively, $f^{XX'}_{l L l'}$ and $h^{XX'}_{l L l'}$  are the lensing/CIP response functions for different quadratic pairs (see Table \ref{tab:CIP_resp}) and are defined in terms of the unmodulated power spectrum, ${C}_l^{XX'}$, and the lensing angular/CIP response functions 
\begin{eqnarray}
_{\pm s} G_{l L l'} &\equiv& \left[L(L+1) + l'(l'+1) - l (l + 1)\right] \sqrt{\frac{(2L+1)(2 l +1)(2 l'+1)}{16\pi}} \left(\begin{array}{ccc}l & L & l' \\ \pm s & 0 & \mp s\end{array}\right),\\
_{\pm s}H_{l L l'} &\equiv& \sqrt{\frac{(2L+1)(2 l +1)(2 l'+1)}{4\pi}} \left(\begin{array}{ccc}l & L & l' \\ \pm s & 0 & \mp s\end{array}\right).
\end{eqnarray}

\begin{table}[!ht]
\begin{tabular*}{11cm}{@{\extracolsep{\fill}}cccc}
\\\noalign{\smallskip}\hline\noalign{\smallskip}
{\rm XX}$'$&$f^{\rm X X'}_{l Ll'}$&$h^{\rm X X'}_{l Ll'}$&$l+l'+L$\\\noalign{\smallskip}\hline\noalign{\smallskip}
TT &$ \left({C}_{l}^{\rm TT}{\,}_0G_{l' L l}+{C}_{l'}^{\rm TT}{\,}_0G_{l L l'}\right)$&$ \left(C_{l}^{\rm T,dT}+{C}_{l'}^{\rm T,dT}\right){\,}_0H_{l' L l}$&even \\\noalign{\smallskip}
TE&$ {C}_{l'}^{\rm TE}{\,}_2G_{l' L l}+{C}_{l}^{\rm TE}{\,}_0F_{l L l'}$&$ C_{l'}^{\rm T,dE}{\,}_2H_{l' L l}+C_{l}^{\rm E,dT}{\,}_0H_{l L l'}$& even  \\ \noalign{\smallskip}
TB&$-i{C}_{l'}^{\rm TE}{\,}_2G_{l' L l} $&$-iC_{l'}^{\rm T,dE}{\,}_2H_{l' L l} $& odd
\\ \noalign{\smallskip}
EE&$ {C}_{l}^{\rm EE}{\,}_2G_{l' L l}+C_{l'}^{\rm EE}{\,}_2G_{l L l'}$&$ \left(C_{l}^{\rm E,dE}+C_{l'}^{\rm E,dE}\right){\,}_2H_{l' L l}$& even  \\ \noalign{\smallskip}
EB &$ i\left[{C}_{l}^{\rm EE}{\,}_2G_{l' L l} -{C}_{l'}^{\rm BB}{\,}_2G_{l L l'}\right]$&$ -iC_{l}^{\rm E,dE}{\,}_2H_{l' L l}$& odd   \\ \noalign{\smallskip}
BB&$ {C}_{l}^{\rm BB}{\,}_2G_{l' L l}+C_{l'}^{\rm BB}{\,}_2G_{l L l'}$& $(C_{l'}^{B,dB}+ C_{l}^{B,dB})H_{l' L l}$ & even  \\\noalign{\smallskip}\hline\noalign{\smallskip}\end{tabular*}
\caption{The lensing and CIP response functions. ``Even'' and ``odd'' indicate that the functions are non-zero only when $L + l + l'$ is even or odd, respectively. Note that the B-mode autocorrelation, BB, vanishes at linear order in the CIP field.} 
\label{tab:CIP_resp}
\end{table}

Planck estimates the lensing potential from observations of the CMB using the formalism presented in Ref.~\cite{Okamoto:2003zw} which 
establishes a minimum-variance estimator for $d_{LM}\equiv\sqrt{L(L+1)}\phi_{LM}$:
\be
\hat d^{\alpha}_{LM} = A_L^\alpha \sum_{l m, l'm'} X_{l m} X'_{l' m'} \wigner l m {l'} {m'} L {-M} F^{XX'}_{l l'L},
\label{eq:dhat}
\ee
where $A_L^\alpha$ and $F^\alpha_{l l'L}$ are
\begin{eqnarray}
A_L^{\alpha} &=& L(L+1)(2L+1)\left\{\sum_{l_1 l_2} g^\alpha_{l_1 L l_2} f^\alpha_{l_1 L l_2}\right\}^{-1},\\
F^{\alpha}_{l Ll'} &\equiv&\frac{C^{XX,{\rm t}}_{l'} C^{X'X',{\rm t}}_l f^{\alpha*}_{l L l'} - (-1)^{l + L + l'} C_{l}^{XX',{\rm t}} C^{XX',{\rm t}}_{l'} f^{\alpha*}_{l' L l}}{C_{l}^{XX,{\rm t}} C_{l'}^{XX,{\rm t}} C_{l}^{X'X',{\rm t}}C_{l'}^{X'X',{\rm t}} - (C_{l}^{XX',{\rm t}}C_{l'}^{XX',{\rm t}})^2},\label{eq:optimal_filter}
\end{eqnarray}
and yield, in the absence of a CIP modulation, an optimal estimator with an expectation value
\be
\hat C_L^{\phi \phi} = \frac{1}{2L+1} \sum_\alpha\sum_{M=-L}^L w_\alpha \hat{d}^\alpha_{LM} \frac{\hat{d}^{*\alpha}_{LM}}{L(L+1)}-B_L,
\ee
where \cite{Okamoto:2003zw}
\begin{eqnarray}
w_\alpha &\equiv& N_L^{\rm mv} \sum_{\beta} ({\bf N}_L^{-1})^{\alpha \beta}, \\
N_L^{\alpha \beta} &\equiv& \frac{A_L^{\alpha *} A_L^\beta}{L(L+1)(2L+1)} \sum_{l_1 l_2} \{ F_{l_1 L l_2}^{\alpha*} [C^{XY,{\rm t}}_{l_1} C^{X'Y',{\rm t}}_{l_2} g^\beta_{l_1Ll_2}+(-1)^{L+l_1+l_2} C^{XY',{\rm t}}_{l_1} C^{X'Y,{\rm t}}_{l_2} F^\beta_{l_2 L l_1}, \quad {\rm and}\\
N_L^{\rm mv} &\equiv& \left[\sum_{\alpha \beta} ({\bf N}_L^{-1})^{\alpha \beta}\right]^{-1}
\end{eqnarray}  
are weights chosen to yield an overall optimal estimator and $B_L$ are the standard Gaussian and non-Gaussian contributions to the CMB four-point correlation \cite{PhysRevD.67.123507,2011PhRvL.107b1301D,Ade:2015zua}.

With a non-zero CIP contribution, the lensing estimator in Eq.~(\ref{eq:dhat}) becomes biased:
\be
\VEV{\hat C_L^{\phi \phi}} =  \phi_{LM} + \sum_{\alpha,\beta} w_{\alpha} w_{\beta}Q^\alpha_LQ^\beta_L \Delta_{LM}, 
\ee
where
\be
Q^\alpha_L \equiv \dfrac{\sum_{l l'} h^\alpha_{l L l'} F^\alpha_{l L l'} }
{\sum_{l l'} f^\alpha_{l L l'} F^\alpha_{l L l'}}.
\ee

\section{Computing the second-order CIP effects}
\label{sec:CIPeffects_secondorder}

We start by noting that in the absence of the CIPs the power spectrum, $C_l^{\rm XX'}$, can be written in terms of an integral over wavenumber \cite{Seljak:1996is}:
\begin{equation}
C_l^{\rm XX'} = \frac{2}{\pi} \int k^2 dk P_{\Phi}(k) X_l(k) X'_l(k),
\label{eq:los}
\end{equation}
where the $X_l(k)$ weighted integrals of the transfer function, $f^X(\eta)$, along the line of sight:
\begin{equation}
X_l(k) \equiv \int d \eta f^X(\eta) j_l[k(\eta_0 - \eta)].
\end{equation}
A CIP causes a modulation of the transfer function, yielding an observed power spectrum \cite{Grin:2011tf}
\begin{equation}
C^{XX', {\rm obs}}_l \simeq C^{XX'}_l \big|_{\Delta = 0} +\sum_{Ll'} C_L^\Delta C^{dX,dX'}_{l'} ({_{s_X}K_{ll'}^L})({_{s_{X'}}K_{ll'}^L})  G_{Ll'} + \frac{1}{2}\Delta_{\rm rms}^2(R_{\rm CMB}) \left(C_l^{X,d^2X'}+C_l^{X',d^2X}\right), \label{eq:grinTTobs}
\end{equation}
where
\begin{eqnarray}
C_l^{dT,dT} &\equiv& \frac{2}{\pi} \int k^2 dk P_{\Phi}(k) \left(\frac{dT_l(k)}{d\Delta}\right)^2,\\
C_l^{T,d^2T} &\equiv& \frac{2}{\pi} \int k^2 dk P_{\Phi}(k) T_l(k) \frac{d^2T_l(k)}{d\Delta^2},\\
G_{L,l'} &\equiv& \frac{(2L+1)(2l'+1)}{4\pi}, \\
_sK_{ll'}^L &\equiv& \left(\begin{array}{ccc}l & L & l' \\s & 0 & -s\end{array}\right),
\end{eqnarray}
and $s_X=0$ for $X=T$ and $s_X = 2$ for $X=E,B$. 

Throughout this work we have taken the CIP modulation to be scale-invariant so that $C_L^\Delta \propto L^{-2}$.  This means 
that the terms in the sum of Eq.~(\ref{eq:modecoupterm}) will be dominated by smaller values of $L$.  Since the variation of the baryon density mainly affects the physics at the surface of last scattering and earlier, the $C^{ dX,dX'}_{l'}$ term is only significant on scales smaller than the angular scale of the horizon at decoupling--so, $l' \gtrsim 100$--this means that within the sum the terms which dominate have $L \ll l'$. The Wigner 3-j symbol in $_sK^L_{ll'}$ is only non-zero when $l$, $L$, and $l'$ satisfy the triangle inequality: 
\begin{equation}
|l-L| \leq l' \leq l+L.
\end{equation}
This set of inequalities applies to the three sides of a triangle.  If $l'$ and $L$ are two sides of a triangle, given that $l'\gg L$, the only way to complete it is to add another side with $l \simeq l'$.  This is the case and it is straight forward to show that the summand in Eq.~(\ref{eq:grinTTobs}) peaks at $l' = l$ and that we have
\begin{equation}
\sum_{l'}({_{s_X}K_{ll'}^L})({_{s_{X'}}K_{ll'}^L}) \simeq \frac{1}{2l'+1}\delta_{l,l'}.
\end{equation}
Therefore we have
\begin{equation}
\sum_{Ll'} C_L^\Delta C^{dX,dX'}_{l'} ({_{s_X}K_{ll'}^L})({_{s_{X'} }K_{ll'}^L})  G_{Ll'} \simeq C_l^{dX,dX'}\sum_L \frac{2L+1}{4\pi} C_L^\Delta = \Delta_{\rm rms}^2(R_{\rm CMB})C_l^{dX,dX'} .\label{eq:modecoupterm}
\end{equation}
Eq.~(\ref{eq:los}) tells us that the second derivative of the \emph{power spectrum} with respect to $\Delta$ can be written 
\begin{equation}
\frac{1}{2}\frac{d^2 C_l^{XX'}}{d\Delta^2}   = \frac{2}{\pi} \int k^2 dk P_{\Phi}(k) \left[\left(\frac{dX_l(k)}{d\Delta}\right)^2 + \frac{1}{2} \left(X'_l(k) \frac{d^2 X_l(k)}{d\Delta^2}+X_l(k) \frac{d^2 X'_l(k)}{d\Delta^2}\right)\right].\label{eq:munozTTobs}
\end{equation}
This shows that in the presence of a scale-invariant CIP power spectrum the full calculation of the $C_l$s in Eq.~(\ref{eq:grinTTobs}) can be replaced with the less computationally intensive expression 
\begin{equation}
C^{\rm XX', {\rm obs}}_l \simeq C^{\rm XX'}_l \big|_{\Delta = 0} +  \frac{1}{2}\Delta_{\rm rms}^2(R_{\rm CMB})\frac{d^2 C_l^{\rm XX'}}{d\Delta^2} \bigg|_{\Delta = 0}.
\label{eq:fullClobs}
\end{equation}
When computing the observed $C_l$s in our MCMC we use an efficient double-sided derivative to numerically calculate the CIP effects.  We detail the accuracy of our numerical derivatives in Appendix \ref{sec:efficient}.  Note that if the CIP spectrum had a blue tilt, or additional power at small scales, the above equation for the observed power spectrum would no longer be a good approximation. 

\section{Efficient computation of the second order effects of CIPs on the CMB power spectrum}
\label{sec:efficient}

Given the number of evaluations of the power spectrum during an MCMC analysis it is essential to use a computationally efficient method to compute the CIP effecs on the CMB power spectrum.  As shown in Eqs.~(\ref{eq:flatClobs}) and (\ref{eq:fullClobs}) in the flat-sky approximation and with the full sky, respectively, the CIP modulated CMB power spectrum can be computed by taking the second derivative of the unmodulated power spectrum with respect to the CIP field $\Delta$.  As discussed in Sec.~\ref{sec:cipsum}, the CIP field is a spatial modulation of the baryon fraction, so this is equivalent to taking the second derivative of the unmodulated CMB power spectrum with respect to $\Omega_b$.  In this Appendix we outline the numerical techniques we developed to efficiently and accurately compute this derivative. 

We compute the numerical derivative of the power spectrum using the finite difference:
\begin{equation}
\frac{\partial^2 C_l}{\partial \Delta^2}  = \frac{C_l[\Delta(1+\epsilon) ] - 2C_l[\Delta] + C_l[ \Delta(1-\epsilon)]}{\epsilon^2 \Delta^2},
\label{twosidederiv}
\end{equation}
where $\epsilon$ is the fractional step in $\Delta$.  In order to determine the value of $\epsilon$ which yields the most accurate derivative we compared Eq.~(\ref{twosidederiv}) to the second derivative computed from a densely sampled polynomial fit to $C_{l}(\Delta)$.  
In particular, we generated 100 evenly spaced samples $C_{l}(\Delta)$ within a range of $-.25 < \Delta < .25$ and fit a 6th order polynomial.  We found that increasing the number of samples and polynomial order beyond these values lead to less than a 0.1\% change in the second order derivative across the entire range of $2 \leq l \leq 5000$. 

\begin{figure}[!ht]
    \centering
    \includegraphics[width=0.6\textwidth]{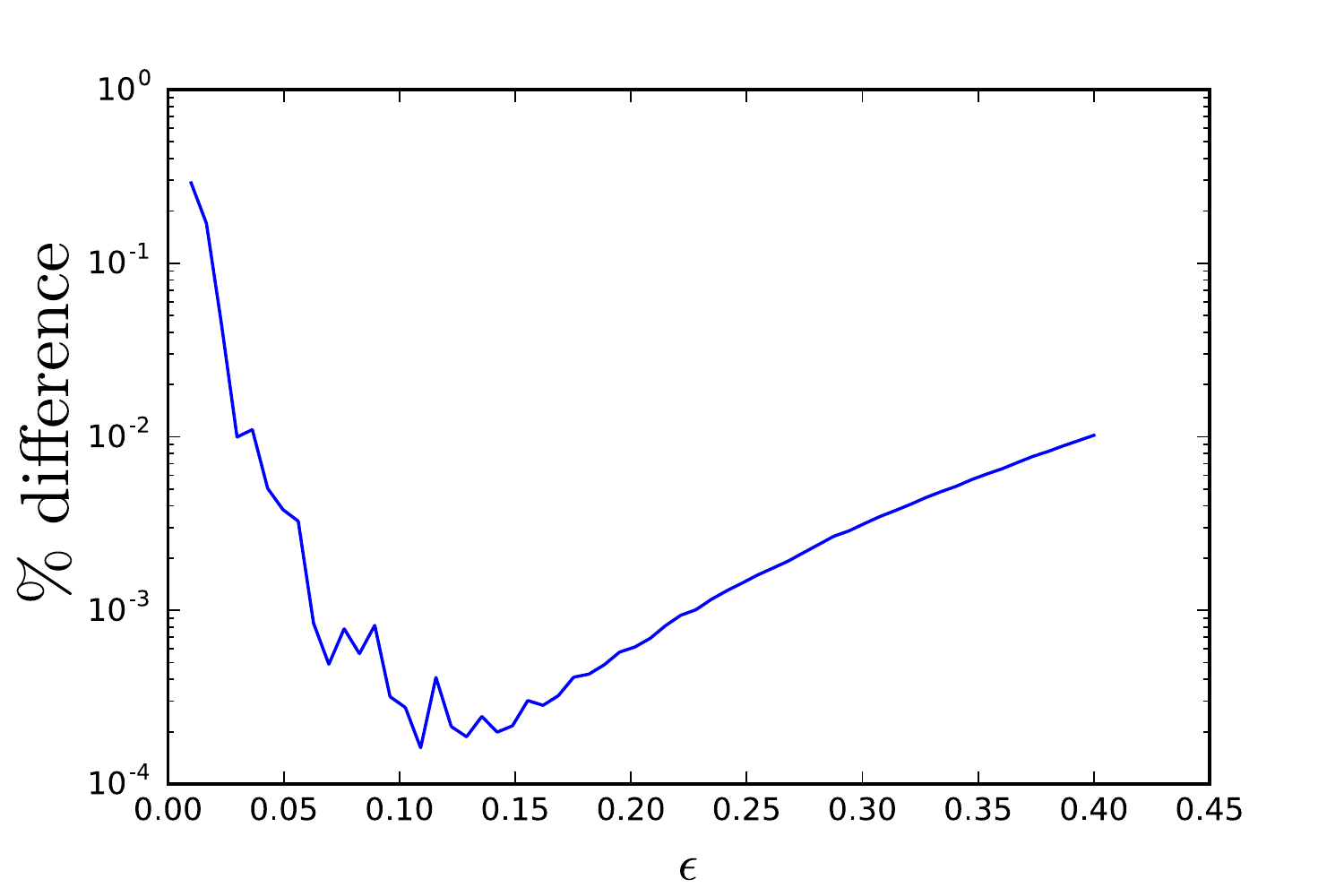}
    \caption{Percent difference between the polynomial derivative method and the finite difference [Eq.~(\ref{twosidederiv})] plotted as a function of the step size $\epsilon$.}
\label{fig:goodness_steps}
\end{figure}

Fig.~\ref{fig:goodness_steps} demonstrates the effect of the size of the step size $\epsilon$ in the finite difference formula on how closely this approximation approaches the polynomial-fitting derivative method. For the accuracy level that we used (in particular we used \texttt{AccuracyBoost = 1} in \texttt{camb}), we can see that the percent difference between our finite difference and polynomial fit derivatives achieves a minimum of less than 0.1\% in the range from $.1 \lesssim \epsilon \lesssim .15$. In the calculations of the CIP modulated CMB power spectrum we used a step size $\epsilon = 0.1$. 

\section{Minimum-variance CIP estimator}
The CIP field $\Delta_{LM}$ induces off-diagonal correlations between CMB multipoles, as shown in Table \ref{tab:CIP_resp} (and derived in Refs. \cite{Grin:2011tf,He:2015msa}). As with lensing, these correlations can be used to derive a minimum variance estimator for CIPs, which for a single pair of observables ($X,X'\in\left\{{\rm T,E,B}\right\}$) is
\begin{align}
\hat{\Delta}_{LM}^{XX'}=& N_{L}^{XX'} \sum_{lm l'm'}X_{lm}X'_{l'
m'}g_{l L l'}^{XX'}(-1)^{m}\sqrt{\frac{(2L+1)(2l+1)(2l'+1)}{4\pi}}	
\wigner{l}{m}{l'}{m'}{L}{M},\label{eq:single_estimator}\\
\left[N_{L}^{XX'}\right]^{-1}=&\sum_{l l'}G_{l l'}h_{l L l'}^{XX'}g_{ l L l'}^{ XX'},\label{eq:single_norm}\\
G_{l l'}\equiv& \frac{(2l+1)(2l'+1)}{4\pi},
\end{align}where the filters $h_{l L l'}^{X X'}$ are as given in Eq.~(\ref{eq:optimal_filter}), but using the CIP response functions in Table \ref{tab:CIP_resp}.

The (Gaussian) reconstruction noise is then
\begin{eqnarray}
\left \langle |\hat{\Delta}_{LM}^{XX'}-\Delta_{LM}|^{2}\right \rangle_{\rm CMB}=N_{L}^{XX'}.\label{eq:variance_general}
\end{eqnarray}
Using different observable pairs, one can obtain the combined  
total minimum-variance estimator
\begin{eqnarray}
\hat{\Delta}_{LM}&=&\sum_{\alpha} w_L^{\alpha}\hat{\Delta}_{LM}^{\alpha},\qquad
w^{\alpha}_L=N_{L}^{\Delta\Delta}\sum_{\beta}\left( \mathcal{M}_{L}^{-1}\right)^{\alpha,\beta},\qquad
 \left[N_{L}^{\Delta \Delta}\right]^{-1} \equiv{\sum_{\alpha\beta}\left(\mathcal{M}_{L}^{-1}\right)^{\alpha,\beta}}. \label{eq:full_estimator}
\end{eqnarray} The estimator covariance-matrix $\mathcal{M}_{L}$ is at every $L$ a rank-$2$ tensor over observable pairs. The indices $\alpha$ and $\beta$ take values over labels for pairs of observables, that is, $\alpha,\beta\in \left\{TT,EE,TE,BT,BE\right\}$. Using 
Eq.~(\ref{eq:single_estimator}), and identities of Wigner coefficients, we obtain an expression for the matrix elements $\mathcal{M}_{L}^{\alpha,\beta}$:\begin{eqnarray}
\mathcal{M}_{L}^{XX', ZZ'}=N_{L}^{XX'}N_{L}^{ZZ'}\sum_{l l'}G_{l l'}g_{L l l'}^{XX'}\left[ {C}_{l'}^{XZ,{\rm t}}{C}_{l}^{X'Z',{\rm t}}g_{L l l'}^{X'Z'*}+\left(-1\right)^{l+l'+L}{C}_{l'}^{XZ',{\rm t}}{C}_{l}^{X'Z,{\rm t}}g_{L l' l}^{ZZ'*} \right]. \label{eq:covmat_total}
\end{eqnarray}
The total-estimator variance is again the inverse normalization factor $N_{L}^{\Delta \Delta}$.  
\label{append:optimal}

\end{appendix}
	
\bibliography{CIP.bib}

\end{document}